\documentclass[11pt]{article}
\usepackage[dvipdfmx]{graphicx}   
\usepackage{amssymb}
\usepackage{array} 
\usepackage[authoryear]{natbib}
\usepackage{bibentry}
\usepackage{amsmath}
\usepackage{float}
\usepackage{verbatim}
\usepackage{color}
\usepackage{bm}
\usepackage{appendix}

\DeclareMathOperator{\tr}{tr}
\DeclareMathOperator{\vecm}{vec}

\DeclareMathOperator{\Normal}{\operatorname{N}}
\DeclareMathOperator{\IG}{\operatorname{IG}}
\DeclareMathOperator{\IW}{\operatorname{IW}}
\DeclareMathOperator{\dBeta}{\operatorname{Beta}}
\DeclareMathOperator{\diag}{diag}
\newcommand{\bvec}[1]{\mbox{\boldmath $#1$}}
\DeclareMathOperator{\boldzero}{\bvec{0}}
\DeclareMathOperator{\boldone}{\bvec{1}}
\DeclareMathOperator{\boldy}{\bvec{y}}
\DeclareMathOperator{\boldV}{\mathbf{V}}
\DeclareMathOperator{\boldeps}{\bvec{\epsilon}}
\DeclareMathOperator{\boldmu}{\bvec{\mu}}
\DeclareMathOperator{\boldphi}{\bvec{\phi}}
\DeclareMathOperator{\boldPhi}{\mathbf{\Phi}}
\DeclareMathOperator{\boldeta}{\bvec{\eta}}
\DeclareMathOperator{\boldrho}{\bvec{\rho}}
\DeclareMathOperator{\boldOmega}{\mathbf{\Omega}}
\DeclareMathOperator{\boldLambda}{\mathbf{\Lambda}}
\DeclareMathOperator{\boldlambda}{\bvec{\lambda}}
\DeclareMathOperator{\boldtheta}{\bvec{\theta}}
\DeclareMathOperator{\bolddelta}{\bvec{\delta}}
\DeclareMathOperator{\boldsigma}{\bvec{\sigma}}
\DeclareMathOperator{\boldSigma}{\mathbf{\Sigma}}
\DeclareMathOperator{\boldPsi}{\mathbf{\Psi}}
\DeclareMathOperator{\boldGamma}{\mathbf{\Gamma}}
\DeclareMathOperator{\boldnu}{\bvec{\nu}}
\DeclareMathOperator{\boldxi}{\bvec{\xi}}
\DeclareMathOperator{\boldh}{\bvec{h}}
\DeclareMathOperator{\bolde}{\bvec{e}}
\DeclareMathOperator{\boldR}{\mathbf{R}}

\DeclareMathOperator{\boldA}{\mathbf{A}}

\DeclareMathOperator{\boldI}{\mathbf{I}}
\DeclareMathOperator{\boldx}{\bvec{x}}
\DeclareMathOperator{\boldX}{\mathbf{X}}
\DeclareMathOperator{\boldw}{\bvec{w}}
\DeclareMathOperator{\boldg}{\bvec{g}}
\DeclareMathOperator{\boldm}{\bvec{m}}
\DeclareMathOperator{\boldM}{\mathbf{M}}
\DeclareMathOperator{\boldB}{\mathbf{B}}
\DeclareMathOperator{\boldz}{\bvec{z}}
\DeclareMathOperator{\boldbeta}{\bvec{\beta}}

\DeclareMathOperator{\E}{E}
\DeclareMathOperator{\Var}{Var}
\DeclareMathOperator{\calF}{\mathcal{F}}
\DeclareMathOperator{\boldP}{\mathbf{P}}
\DeclareMathOperator{\boldQ}{\mathbf{Q}}

\newcommand{\dtm}[1]{\lvert#1\rvert}
\newcommand{\abs}[1]{\lvert#1\rvert}
\newcommand{\inv}[1]{\frac{1}{#1}}

\begin{document}
\title{Multivariate Stochastic Volatility Model with Realized Volatilities and Pairwise Realized Correlations}
\author{Yuta Yamauchi\thanks{Graduate School of Economics, The University of Tokyo, Tokyo, Japan. E-mail:mchyuta@gmail.com.} \ and Yasuhiro Omori\thanks{Faculty of Economics, The University of Tokyo, Tokyo, Japan. E-mail:omori@e.u-tokyo.ac.jp. Phone: +81-3-5841-5516. Fax: +81-3-5841-5521.}}
\maketitle
\begin{abstract}
Although stochastic volatility and GARCH (generalized autoregressive conditional heteroscedasticity)  models have successfully described the volatility dynamics of univariate asset returns, extending them to the multivariate models with dynamic correlations has been difficult due to several major problems. First, there are too many parameters to estimate if available data are only daily returns, which results in unstable estimates. One solution to this problem is to incorporate additional observations based on intraday asset returns, such as realized covariances.
Second, since multivariate asset returns are not synchronously traded, we have to use the largest time intervals such that all asset returns are observed in order to compute the realized covariance matrices. However, in this study, we fail to make full use of the available intraday informations when there are less frequently traded assets.
Third, it is not straightforward to guarantee that the estimated (and the realized) covariance matrices are positive definite. 
%


Our contributions are the following: (1) we obtain the stable parameter estimates for the dynamic correlation models using the realized measures, (2) we make full use of intraday informations by using  pairwise realized correlations, (3) the covariance matrices are guaranteed to be positive definite, (4) we avoid the arbitrariness of the ordering of  asset returns, (5) we propose the flexible correlation structure model (e.g., such as setting some correlations to be zero if necessary), and (6) the parsimonious specification for the leverage effect is proposed.
Our proposed models are applied to the daily returns of nine U.S. stocks with their realized volatilities and pairwise realized correlations and are shown to outperform the existing models with respect to  portfolio performances.
\end{abstract}

\clearpage

\section{Introduction}

Modelling the time-varying volatility and the correlations of multivariate time series is one of the most important problems in financial risk management, and there are numerous studies that model the time-varying volatility of univariate time series using the GARCH or stochastic volatility (SV) models.
However, the extension of their models to multivariate model with dynamic correlations has not been straightforward due to the following several major problems.%

First, there are too many parameters to estimate if the only available data are daily returns, which results in unstable estimates.  An intuitive solution to reduce the number of parameters is to introduce the factor structure assuming that a small number of common factors describe the dynamics of time-varying covariance matrices as discussed in the factor stochastic volatility models (e.g. \cite{pitt1999time}, \cite{chib2006analysis} and \cite{lopes2007factor}). However, factor modelling requires the a priori selection of the number of factors and we need to restrict the structure of the factors in order to identify the parameters (e.g. \cite{lopes2004bayesian}). Furthermore, the estimation results and the predictive performance of the model are usually subject to the ordering of the asset returns.

An alternative effective approach is to incorporate additional observations based on the intraday asset returns, such as the realized covariances, which have recently become available in financial markets. In univariate SV models, the realized SV (RSV) models that estimate the time-varying volatilities using the daily returns and realized volatility simultaneously have been proposed to achieve more accurate parameter estimates than those of the SV models using only daily returns and the RSV models outperform SV models in forecasting volatilities (e.g. \cite{TakahashiOmoriWatanabe(09)}, \cite{DobrevSzerszen(10)}, \cite{KoopmanScharth(13)}, \cite{ZhengSong(14)}, 
 \cite{TakahashiWatanabeOmori(16)}). Although the realized volatilities are subject to microstructure noises and nontrading hours and hence are biased estimates of the integrated volatilities, such biases are automatically adjusted within the proposed model. Similarly, the univariate GARCH model is extended to the realized GARCH models which incorporates the realized volatilities into the variance equations and it is shown to lead to substantial improvements in the empirical fit and quantile forecasts over the standard GARCH model that only uses daily returns (\cite{HansenHuangShek(12)}).
%

The extension to the multivariate RSV model is also considered in the Cholesky RSV model (\cite{ShirotaOmoriLopesPiao(17)}). In this model, the Cholesky decompositions of the realized covariance matrices are used as additional sources for measurement equations, and it models the dynamics of the logarithm of the diagonal elements and the off-diagonal elements of Cholesky decomposed covariance matrices respectively. It is shown that the portfolio performances of the proposed model outperformed other SV models without realized measures in the empirical studies, but it should also be noted that the performance of the Cholesky RSV models may depend on the ordering of the asset returns in the vector of the response.

Second, high-frequency data are not always observed at the same time points, which causes difficulties in the extension of the univariate RSV model to the multivariate RSV model. For example, in the Cholesky RSV model, it is implicitly assumed that the all multivariate assets are traded every few minutes when computing the realized covariance matrices. If the multivariate assets are not traded synchronously, we have to use the largest time intervals so that all asset returns are observed when computing the realized covariance matrices. This nonsynchronous trading leads us to ignore some of the frequently traded asset return data, and hence we would fail to make full use of the available intraday informations when there are less frequently traded assets.

Third, it is not straightforward to guarantee that the estimated (and the realized) covariance matrices are positive definite. The model parameters may be difficult to estimate in practice under the constraints that satisfy the positive definiteness. Using the Cholesky decomposition of the time-varying covariance matrices is one way to guarantee the positive definiteness (\cite{ShirotaOmoriLopesPiao(17)}), but it also requires that the multivariate assets are traded synchronously in order to compute the realized covariance matrices as mentioned above. Additionally, the interpretation of each latent variable of the decomposition is not straightforward since it does not correspond to each pair of asset returns and it is subject to the ordering of the asset returns. If we use each element of the realized covariances for each pair of asset returns, it may result in the nonpositive definite covariance matrices. 

To overcome these difficulties, we propose a multivariate realized SV (MRSV) model with pairwise realized correlations, in which we incorporate the dynamic latent correlation variables in addition to latent volatility variables with realized measures for each pairwise correlation and the volatilities in the framework of multivariate SV models with realized volatilities. The model parameters are estimated using Markov chain Monte Carlo simulations, and we sample the latent correlation variables one at a time given the others so that we keep the covariance matrices positive definite. %
The realized Beta GARCH model proposed by \cite{HansenLundeVoev(14)} is a promising multivariate GARCH model with realized measures for volatilities and co-volatilities in which they used measurement equations for the pairwise realized correlations with market returns and modelled dynamics of the Fisher transformed conditional correlation coefficients. However, they focused on the pairwise correlations between the market return and an individual asset return, assuming that the individual asset returns are conditionally independent given the market return. Another useful approach for the joint modelling of returns and realized covariances are based on Wishart processes (e.g. \cite{JinMaheu(13)}, \cite{WindleCarvalho(14)}, \cite{JinMaheu(16)}, \cite{SoLiAsaiJiang(16)}). The covariance matrix is assumed to follow a Wishart distribution whose scale matrix depends on the past realized covariance matrices which are computed using larger time intervals than necessary in order for it to be positive definite.

Our approach, on the other hand, is based on simultaneously modeling the individual volatilities and pairwise covariances, rather than the covariance matrix, and we are able to make full use of the available intraday information, even when there are less frequently traded assets. This finding implies that our model still can be constructed even though some of the realized measures are missing. Additionally, our model is far more flexible in the sense that it is possible to restrict any correlation coefficients to be zero for very high dimensional asset returns data, which reduces the number of parameters and may improve the forecasting performances. Among the multivariate SV models in the literature, our model is a natural extension of a univariate RSV model to a multivariate model, and it gives us a straightforward interpretation of the estimated parameters.

Furthermore, we extend our model to incorporate the leverage effect, which is well-known to exist in stock markets.  The leverage effect refers to the negative correlation between an asset return and its volatility. In other words, a decrease in the stock return is followed by an increase in its volatility. In forecasting the means and covariances of asset returns for portfolio optimization, it is expected that incorporating the leverage effect in econometric models improves the predictive accuracy. However, it may increase the number of parameters that are to be estimated and the realized measures are not available for such an effect. Thus we also consider the parsimonious parameterization for the leverage effect.


%
Our contributions are  as follows: (1) we obtain the stable parameter estimates for the dynamic correlation models using the realized measures, (2) we make full use of the intraday informations by using  pairwise realized correlations, (3) the estimated covariance matrices are guaranteed to be positive definite, (4) we avoid the arbitrariness of the ordering of  asset returns, (5) we propose the flexible correlation structure model and set some correlations to be zero if necessary, and (6) we introduce the parsimonious specification for the leverage effect.

The structure of this paper is as follows. Section 2 introduces the multivariate realized SV model with daily returns, realized volatilities, and pairwise realized correlations. Section 3 describes the estimation algorithms using the Markov Chain Monte Carlo simulation. Section 4 extends it to incorporate the leverage effect. Finally, in Section 5, the proposed model is applied to nine U.S. stock return data and the model with the leverage effect is shown to outperform other competing models with regard to the portfolio performances.

\section{Multivariate realized stochastic volatility model}

This section introduces the multivariate realized  stochastic volatility (MRSV) model, which uses realized measures for the volatility and pairwise correlations of asset returns. By using the additional information of the realized measure for asset returns, we can overcome the curse of dimensionality when estimating the dynamic covariance matrices. 
The Cholesky RSV model proposed by \cite{ShirotaOmoriLopesPiao(17)} also  uses the realized measure of variances and covariances (which we call the realized covariance matrix) in order to estimate the latent covariance matrix of asset returns. However, the realized covariance matrix is less informative when there are less frequent asset returns. This finding is observed because  we require the synchronous observations of all asset return series in order to compute the realized covariance matrix.
In order to utilize the full information of the realized measure for the correlations, we propose using the realized measures for the latent pairwise correlations. It should be noted that the pairwise correlation can be computed if that pair of series is synchronously observed. We call the realized measure for the correlation coefficient the pairwise realized correlation.
Using the pairwise realized correlations, in order to guarantee the positive definiteness of the latent covariance matrices, we propose the MCMC algorithm in which we sample latent correlation coefficients from the conditional posterior distribution so that the matrices are positive definite. 
\subsection{Multivariate stochastic volatility  model with dynamic correlations}
First, we define the multivariate SV (MSV) model without the realized measures. Let $\boldy_t= (y_{1t},\ldots,y_{pt})'$ and $\boldh_t = (h_{1t},\ldots,h_{pt})'$ denote a $p \times 1$ stock return vector and its corresponding log volatility latent vector at time $t$. The basic MSV model is given by
\begin{align}
\label{eq:basicMSV-1}
	&\boldy_t = \boldm_t + \boldV_t^{1/2}\boldeps_t, 
	\quad \boldeps_t \sim \Normal(\boldzero, \boldR_t), \quad 
	\quad t=1,\ldots,T,\\
	&\boldh_{t+1} = \boldmu + \boldPhi (\boldh_t - \boldmu) + \boldeta_t,
	\quad \boldeta_t \sim \Normal(\boldzero, \boldOmega),
	\quad t=1,\ldots,T-1, \\
	&\boldm_{t+1} = \boldm_t + \boldnu_t,  \quad \boldnu_t \sim \Normal(\boldzero, \boldSigma_m), \quad t=1,\ldots,T-1,\\
	&\hspace{3mm} \boldh_1 \sim \Normal(\boldmu,\boldOmega_0),
	 \hspace{3mm} \boldm_1 \sim \Normal(\boldzero, \kappa\boldSigma_m),
\label{eq:basicMSV-4}
\end{align}
where $\boldR_t=\{\rho_{ij,t}\}$ is a correlation matrix, $\boldeps_t = (\epsilon_{1t},\ldots,\epsilon_{pt})'$,
	$\boldeta_t = (\eta_{1t},\ldots,\eta_{pt})'$, and
\begin{align*}
	& 
	\boldnu_t = (\nu_{1t},\ldots,\nu_{pt})', 
	\quad \boldV_t = \diag(\exp(h_{1t}),\ldots,\exp(h_{pt})),\\
	&\boldSigma_m = \diag(\bm{\sigma}_m^{2}),
	\quad \bm{\sigma}_m^2=(\sigma^2_{m,1},\ldots,\sigma^2_{m,p})',\quad
	\boldPhi = \diag(\bm{\phi}),
	\quad \bm{\phi} = (\phi_1,\ldots,\phi_p)'.
\end{align*}
We assume that $h_{it}$ follows a stationary autoregressive process (with its coefficient $\abs{\phi_j} < 1)$ and that the mean process $\bm{m}_t= (m_{1t},\ldots,m_{pt})'$  follows a random walk process.
We denote a diagonal matrix $\mathbf{A}$ with diagonal elements $\bm{a}=(a_{11},\ldots,a_{mm})'$ as $\mathbf{A}=\mbox{diag}(\bm{a})$.
For the initial distributions of $\bm{m}_1$ and $\bm{h}_1$, we set $\kappa$ to some large constant  for $\bm{m}_1$ for simplicity and  set $\boldOmega_0$ to satisfy the stationary condition $\boldOmega_0 = \boldPhi \boldOmega_0 \boldPhi + \boldOmega$ for $\bm{h}_1$ such that
\begin{align}
\label{eq:initial_dist}
	\vecm(\boldOmega_0) = (\boldI_{p^2} - \boldPhi \otimes \boldPhi)^{-1} \vecm(\boldOmega),
\end{align}
where $\mathbf{I}_{p^2}$ denotes a $p^2\times p^2$ unit matrix.
In order to model the dynamics of the correlation matrix,
we consider the following Fisher transformation $g_{ij,t+1}$ of the correlation coefficient $\rho_{ij,t}$, and assume that it follows  a random walk process for simplicity:
\begin{align}
	& g_{ij,t+1}=g_{ij,t}+\zeta_{ij,t} , \qquad \zeta_{ij,t}\sim \text{i.i.d. }\Normal(0,\sigma^2_{\zeta,ij}), \
	\quad t=1,\ldots,T-1, \\
\label{eq:basicMSV-6}
	& g_{ij,1} \sim \Normal(0,\kappa \sigma^2_{\zeta,ij}), 
	\hspace{9mm} g_{ij,t} = \log(1 + \rho_{ij,t}) - \log(1 - \rho_{ij,t}),
\end{align}
for $i,j=1,\ldots, p$ $(j<i)$ and we denote
$\bm{\rho}_t   =   (\rho_{21,t},\ldots,\rho_{p\hspace{0.1mm}p-1, t})'$,
$\bm{g}_t  =  (g_{21,t},\ldots,g_{p\hspace{0.1mm}p-1, t})'$, 
$\bm{\zeta}_t   =  (\zeta_{21,t},\ldots,\zeta_{p\hspace{0.1mm}p-1, t})'$,
and
$\bm{\sigma}^2_{\zeta}   =   (\sigma^2_{\zeta,21},\ldots,\sigma^2_{\zeta, p\hspace{0.1mm}p-1})'$.
\\
\noindent
{\it Non-arbitrary ordering of asset returns and the flexible correlation structure.} 
We note that above specifications (\ref{eq:basicMSV-1}) -- (\ref{eq:basicMSV-6}) are independent of the ordering of the asset returns in $\bm{y}_t$, while the conventional factor SV models or the Cholesky SV models (\cite{ShirotaOmoriLopesPiao(17)}) may be affected by the ordering. Further, it allows us to model the structure of the correlations in a flexible way. For example, we can easily restrict some correlation coefficients to be zero when the dimension of $\bm{y}_t$ is very high.\\

\noindent
{\it Remark} 1.  It is easy to assume that $\bm{m}_t$ and $g_{ij,t}$ follow stationary autoregressive processes. However, since it imposes the mean reversion properties on these processes, we would rather consider random walk processes without such properties for simplicity. For the long term prediction, we may need such a stationarity condition.
\subsection{Realized stochastic volatilities and pairwise realized correlations}
{\it Realized measures as an additional source of information}. In the above MSV models, there are too many parameters to estimate using only daily asset returns, and the parameter estimates are often unstable. Recently, high frequency data in the financial markets have become available, and they play a more important role in the finance-related empirical studies, since the realized measures of the variances and covariances, are more informative estimators of the true variances and covariances (see e.g. \cite{AndersenBollerslevDieboldLabys(01)}, \cite{AndersenBollerslevDieboldEbens(01)}, \cite{BarndorffShephard(02)}, \cite{BarndorffShephard(04)}).

 Let $x_{it}= \log RV_{it}$ and $w_{ij,t} = \log \{(1 + RCOR_{ij,t})/(1 - RCOR_{ij,t})\}$ where $RV_{it}$ and $RCOR_{ij,t}$ are the realized measures of the volatility of the $i$-th asset return and the correlation between $i$-th and $j$-th asset returns at time $t$. Thus we introduce the following additional measurement equations based on the realized measures:
\begin{align}
	& x_{it}=\xi_{i}+h_{it}+u_{it} \qquad u_{it} \sim \Normal(0,\sigma^2_{u,i}), \quad t=1,\ldots,T, \\ 
\label{eq:basicMSV-8}
	& w_{ij,t}=\delta_{ij}+g_{ij,t}+v_{ij,t}
	\qquad v_{ij,t} \sim \Normal(0,\sigma^2_{v,ij}), \quad t=1,\ldots,T,
\end{align}
for $i,j=1,\ldots, p$ $(i>j)$. The terms $\xi_{j}$ and $\delta_{ij}$ are included in order to adjust the biases due to the microstructure noise, nontrading hours, nonsynchronous trading and so forth. The multivariate realized stochastic volatility model with pairwise realized correlations is defined by  (\ref{eq:basicMSV-1}) -- (\ref{eq:basicMSV-8}).
We denote
$\bm{x}_t = (x_{1t},\ldots,x_{pt})', \quad \bm{w}_t = (w_{21,t},\ldots,w_{p\hspace{0.1mm}p-1,t})'$,
$\boldxi = (\xi_1,\ldots,\xi_p)'$, 
$\bolddelta = (\delta_{21},\ldots,\delta_{p\hspace{0.1mm}p-1})'$,
$\bm{u}_t = (u_{1t},\ldots,u_{pt})'$, 
$\bm{v}_t = (v_{21,t},\ldots,v_{p\hspace{0.1mm}p-1,t})'$,
$\bm{\sigma}^2_u = (\sigma^2_{u,1},\ldots,\sigma^2_{u,p})'$, 
and $\bm{\sigma}^2_v = (\sigma^2_{v,21},\ldots,\sigma^2_{v,p\hspace{0.1mm}p-1})'$. \\
\noindent
{\it Use of pairwise realized correlations.}
Given the realized correlation $RCOR_{ij,t}$, we will use the pairwise realized correlations.
If there is less frequent series of asset returns, the realized covariance matrix may lose a large part of the information since it is calculated only when all the series are synchronously observed. On the other hand, the pairwise realized correlation coefficients can be respectively calculated for each pair of series of returns; therefore, we can use the full information of the realized measures for the correlations.
Moreover, we can estimate the parameters even if we cannot obtain the realized measures for some pairs. \\
\noindent
{\it Bias corrections of the realized measures}.
The realized volatilities and pairwise realized correlations have more information about the true volatilities and correlations, but there may be biases due to the market microstructure noise, nontrading hours, nonsynchronous trading and so forth. In order to correct these biases in the realized measures, we model the observation equations of the realized volatilities and pairwise realized correlations with bias adjustment terms, $\xi_j$ and $\delta_{ij}$. Although daily returns have relatively less information about the true volatilities and correlations, they are less subject to the biases that are caused by the high frequency data. Therefore, we can estimate the biases in the realized measures using the information of daily returns and also get additional information with regard to the true volatilities and correlations using the realized measures.

\section{Markov chain Monte Carlo estimation}
\subsection{Prior distributions for parameters}
Since there are many latent variables in our proposed model and hence it is difficult to evaluate the likelihood, we take the Bayesian approach and estimate the model parameters using the Markov chain Monte Carlo simulation. First we assume the prior distribution of $\boldtheta\equiv (\bm{\phi},\boldmu,\boldxi,\bolddelta, \boldsigma_{u}^2,\boldsigma_{v}^2,\boldsigma_{\zeta}^2,\mathbf{\Sigma}_m,\boldOmega)$ as follows.
For the prior distributions of $\mu_i, \xi_i$ and $\delta_{ij}$, we assume multivariate independent normal distributions.
The prior distributions of $\sigma_{u,i}^2,\sigma_{v,ij}^2,\sigma_{\zeta,ij}^2$ and $\sigma_{m,i}^2$ are  assumed to be independent inverse gamma distributions. For $\phi_i$ and $\boldOmega$, we assume $(1+\phi_i)/2 \sim \dBeta(a,b)$ and an inverse Wishart distribution respectively. In summary, we assume the following prior distributions:
\begin{align}
	& \mu_i \sim \Normal(m_\mu,s^2_{\mu}), \quad  \xi_i \sim \Normal(m_{\xi},s_{\xi}^2), \quad \delta_{ij} \sim \Normal(m_{\delta},s_{\delta}^2), \\
	& \sigma^2_{u,i} \sim \IG\left(\frac{n_u}{2},\frac{d_u}{2}\right), \quad
	\sigma^2_{v,ij} \sim \IG\left(\frac{n_v}{2},\frac{d_v}{2}\right), 
	\quad \sigma^2_{\zeta,ij} \sim \IG\left(\frac{n_\zeta}{2},\frac{d_\zeta}{2}\right), \\
	& \frac{1+\phi_i}{2} \sim \dBeta(a,b),  \quad \sigma^2_{m,i} \sim \IG\left(\frac{n_m}{2},\frac{d_m}{2}\right), 
	\quad \Omega \sim \IW(\nu,\bvec{S}), 
\end{align}
for $i, j=1,\ldots,p$ $(j<i)$, and $a, b,m_\mu,s_{\mu}, m_{\xi},s_{\xi}, m_{\delta},s_{\delta}, n_u, d_u, n_v, d_v,n_{\zeta}, d_{\zeta}, n_m, d_m, \nu,\bvec{S}$ are hyperparameters. 
\\
\noindent
{\it Remark} 2. The particle MCMC may be a possible alternative estimation method to the MCMC below for the univariate models, but it may not be appropriate for the multivariate models since the discrete approximation to the high dimensional state distribution often results in the degeneracy of the particles.
\subsection{Markov chain Monte Carlo algorithm}
Let $\bm{g}=(\bm{g}_1',\ldots,\bm{g}_T')'$, $\bm{h}=(\bm{h}_1',\ldots,\bm{h}_T')'$ and $\bm{m}=(\bm{m}_1',\ldots,\bm{m}_T')'$. Further, let   $\bm{w}=(\bm{w}_1',\ldots,\bm{w}_T')'$, $\bm{x}=(\bm{x}_1',\ldots,\bm{x}_T')'$ and $\bm{y}=(\bm{y}_1',\ldots,\bm{y}_T')'$. 
In order to conduct the statistical analysis of the parameters, we implement the Markov chain Monte Carlo simulation in nine blocks. The MCMC sampling algorithm is described in more details in the following subsections.
Let $\boldtheta_{\backslash \boldbeta}$ denote the parameter $\boldtheta$ excluding $\boldbeta$. Then,
\begin{enumerate}
\item
Initialize $\boldg, \boldh, \boldm$ and $\boldtheta$.
\item
Generate $\boldg \vert \boldtheta,\boldh,\boldm,\boldw,\boldx,\boldy$.
\item
Generate $\boldh \vert \boldtheta,\boldm,\boldg,\boldw,\boldx,\boldy$.
\item
Generate $\boldm \vert \boldtheta,\boldh,\boldg,\boldw,\boldx,\boldy$.
\item
Generate $\boldphi \vert \boldtheta_{\backslash\boldphi}, \boldh,\boldm,\boldg,\boldw,\boldx,\boldy$.
\item
Generate $(\boldmu,\boldxi, \bolddelta) \vert \boldtheta_{\backslash(\boldmu,\boldxi,\bolddelta,)}, \boldh,\boldm,\boldg,\boldw,\boldx,\boldy$.
\item
Generate $(\boldsigma_{u}^2,\boldsigma_{v}^2,\boldsigma_{\zeta}^2,\boldSigma_m) \vert \boldtheta_{\backslash(\boldsigma_{u}^2,\boldsigma_{v}^2,\boldsigma_{\zeta}^2,\boldSigma_m)}, \boldh,\boldm,\boldg,\boldw,\boldx,\boldy$.
\item
Generate $\boldOmega\vert \boldtheta_{\backslash\boldOmega}, \boldh,\boldm,\boldg,\boldw,\boldx,\boldy$.
\item Go to Step 2.
\end{enumerate}
\subsubsection{Generation of $\bm{g}_t$ for the dynamic correlation matrix $\mathbf{R}_t$}
\label{sec:mcmc_gt}
The conditional posterior probability density function of $g_{ij,t}$ given other parameters and latent variables is
\begin{align}
&\pi(g_{ij,t} \vert \cdot)
	\propto
	\exp\left \{-\inv{2 \sigma_{t*}^2} (g_{ij,t} - m_{t*})^2 +r(g_{ij,t})\right\},
\\
&\label{eq:r_g}
\hspace{8mm}	r(g_{ij,t})
	=
	-\frac{1}{2}\log\dtm{\boldR_t}  -\frac{1}{2}(\boldy_t - \boldm_t)'(\boldV_t^{1/2}\boldR_t\boldV_t^{1/2})^{-1} (\boldy_t - \boldm_t),
\end{align}
where
\begin{eqnarray}
m_{t*} & = & \left\{
	\begin{array}{ll}
	\sigma_{t*}^2 \left\{\sigma^{-2}_{\zeta,ij} g_{ij,2} + \sigma^{-2}_{v,ij} (w_{ij,1} - \delta_{ij})\right\}, & t=1,\\
	\sigma_{t*}^2 \left\{\sigma^{-2}_{\zeta,ij} (g_{ij,t-1}+g_{ij,t+1}) + \sigma^{-2}_{v,ij} (w_{ij,t} - \delta_{ij})\right\}, & t=2,\ldots,T-1,\\
	\sigma_{t*}^2 \left\{\sigma^{-2}_{\zeta,ij} g_{ij,T-1} + \sigma^{-2}_{v,ij} (w_{ij,T} - \delta_{ij})\right\}, & t=T,
	\end{array}
\right.
\end{eqnarray}
and
\begin{eqnarray}
\sigma_{t*}^2&=&\left\{
	\begin{array}{ll}
	\left\{(\kappa^{-1}+1)\sigma^{-2}_{\zeta,ij} + \sigma^{-2}_{v,ij}\right\}^{-1}, & t=1,\\
	\left(2\sigma^{-2}_{\zeta,ij} + \sigma^{-2}_{v,ij}\right)^{-1}, & t=2,\ldots,T-1,\\
	\left(\sigma^{-2}_{\zeta,ij} + \sigma^{-2}_{v,ij}\right)^{-1}, & t=T.
	\end{array}
\right.
\end{eqnarray}
\noindent
{\it Positive definiteness of $\mathbf{R}_t$}. We use an identity matrix  for the initial value of $\mathbf{R}_t$  when implementing the MCMC. Thus, given the current correlation matrix $\mathbf{R}_t$, we generate each correlation coefficient $\rho_{ij,t}$ (or equivalently $g_{ij,t})$ so that we guarantee that the proposed $\mathbf{R}_t^*$ is the correlation matrix. We first state the condition for  $\rho_{ij,t}$ to guarantee that the proposed $\mathbf{R}_t^{*}$ is positive definite given the other elements of $\boldR_t$ and other  $\rho_{ij,s}$ $(s\neq t)$ .\\

\noindent
{\bf Proposition 1}.
\hspace{1mm}
Suppose that $\boldR_t=\{\rho_{ij,t}\}$ is a correlation matrix and let $\boldrho_{it}$ denote the transpose of the $i$-th row vector of $\boldR_t$ excluding 1, $\boldrho_{it} = (\rho_{i1,t},\ldots,\rho_{i\hspace{0.1mm}i-1,t},\rho_{i\hspace{0.1mm}i+1,t},\ldots,\rho_{ip,t})'$, and $\boldR_{it}$ denotes the submatrix excluding the $i$-th row and the $i$-th column from $\boldR_t$.
The condition for $\rho_{ij,t}$ to guarantee that $\mathbf{R}_t$ is positive definite is 
$\rho_{ij,t} \in (L_{ijt}, U_{ijt})$ where bounds $L_{ijt}$ and $U_{ijt}$ are given by
\begin{align}
\label{eq:pd}
	\frac{-\bvec{b}_j'\boldrho_{i,-j,t} \pm \sqrt{(\bvec{b}_j'\boldrho_{i,-j,t})^2-a_j(\boldrho'_{i,-j,t}\mathbf{C}_j\boldrho_{i,-j,t}-1)}}{a_j},
\end{align}
and $\boldrho_{i,-j,t}$ is the vector excluding the $j$-th element of $\boldrho_{it}$, $a_j$ is the $(j,j)$-th element of $\boldR_{it}^{-1}$, $\bvec{b}_j$ is the vector excluding $a_j$ from the $j$-th column of $\boldR_{it}^{-1}$, and $\mathbf{C}_j$ is the matrix excluding the $j$-th row and $j$-th column from $\mathbf{R}_{it}^{-1}$.
\\
\noindent
{\bf Proof:} See Appendix \ref{sec:proof1}
\\

Thus we propose a candidate $g_{ij,t}^{\dagger}$ from normal distribution truncated on the interval $(a_{ijt}, b_{ijt}) $, $TN_{(a_{ijt},b_{ijt})}(m_{t*},\sigma_{t*}^2)$, and accept it with probability $\min\{1, \exp(r(g_{ij,t}^{\dagger})-r(g_{ij,t}))\}$,
where
\begin{align}
\label{eq:rho_ij}
	(a_{ijt}, b_{ijt}) \equiv \left(\log\frac{1+L_{ij,t}}{1-L_{ij,t}},\log\frac{1+U_{ij,t}}{1-U_{ij,t}}\right).
\end{align}

\subsubsection{Generation of $\boldh_t$ for the dynamic volatility $\mathbf{V}_t$}
We use a single-move sampler for $\bm{h}_t$ in which we sample $\bm{h}_t$ given the other parameters and latent variables. Such a sampler is efficient when the realized measures are available as the additional information source for $\bm{h}_t$. 
The conditional posterior probability density function of $\boldh_t$ is given by
\begin{align}
&\pi(\boldh_t \vert \cdot)
	\propto
	\exp\left[
	-\frac{1}{2}(\boldh_t - \bm{m}_{t*})'\boldOmega_{t*}^{-1}(\boldh_t - \bm{m}_{t*}) +l(\boldh_t)
	\right],
\\
&
\hspace{8mm}
l(\boldh_t) = -\frac{1}{2}(\boldy_t - \boldm_t)'(\boldV_t^{1/2}\boldR_t\boldV_t^{1/2})^{-1} (\boldy_t - \boldm_t),
\end{align}
where
\begin{eqnarray}
\bm{m}_{t*} & = & \left\{
	\begin{array}{ll}
	\mathbf{\Omega}_{1*}
	\left[
		\boldOmega_0^{-1} \boldmu
		+ \boldPhi \boldOmega^{-1}
		\left\{
			\boldh_{2} - (\boldI_p - \boldPhi)\boldmu
		\right\}
	\right. &
	\\
	\left. \hspace{4.9cm}
			+ \boldSigma_u^{-1}
			(\boldx_1 - \boldxi)
			- \frac{1}{2} \boldone_p
	\right], & t=1,\\
	\boldOmega_{t*}
	\left[
		\boldOmega^{-1} 
		\left\{
		(\boldI_p - \boldPhi)\boldmu + \boldPhi \boldh_{t-1}
		\right\}
	\right. &   \\
	\left.	\hspace{6mm}
		+ \boldPhi \boldOmega^{-1}
		\left\{
			\boldh_{t+1} - (\boldI_p - \boldPhi)\boldmu
		\right\}
		+ \boldSigma_u^{-1}(\boldx_t - \boldxi) - \frac{1}{2} \boldone_p
	\right], & t=2,\ldots,T-1,\\
	\boldOmega_{T*}
	\left[
		\boldOmega^{-1} 
		\left\{	(\boldI_p - \boldPhi)\boldmu + \boldPhi \boldh_{T-1}
		\right\}
		+ \boldSigma_u^{-1}(\boldx_T - \boldxi)- \frac{1}{2} \boldone_p
	\right], & t=T,
	\end{array}
\right.
\\
\mathbf{\Omega}_{t*} & = & \left\{
	\begin{array}{ll}
	\left[
		\boldOmega_0^{-1}
		+ \boldPhi \boldOmega^{-1} \boldPhi
		+ \boldSigma_u ^{-1}
	\right]^{-1}, & t=1,\\
	\left[
		\boldOmega^{-1} + \boldPhi \boldOmega^{-1} \boldPhi + \boldSigma_u ^{-1}
	\right]^{-1}, & t=2,\ldots,T-1,\\
	\left[
		\boldOmega^{-1} + \boldSigma_u ^{-1}
	\right]^{-1}, & t=T,
	\end{array}
\right.
\end{eqnarray}
where $\bm{1}_p$ denotes a $p\times 1$ vector with all elements equal to one.
Therefore, we generate a candidate $\bm{h}_t^{\dagger}$ from $\Normal(\bm{m}_{t*}, \boldOmega_{t*})$, and accept it with probability $\min\{1,\exp(l(\bm{h}_t^{\dagger}) - l(\bm{h}_t))\}$.
See Appendix \ref{sec:mcmc_wo_leverage} for the generations of $\bvec{\theta}$ and $\boldm_t$.

\section{Extension to incorporate the leverage effect}
This section extends our model in order to incorporate the leverage effect. The leverage effect, which corresponds to the well-known negative correlation between asset returns and their volatilities in the stock market, is expected to improve the performance of the forecast of the mean processes and volatility processes of asset returns.
\subsection{Matrix variate normal distribution}
We first define the matrix variate normal distribution and show its probability density function, which will be used in modelling the leverage effect.
\\

\noindent
{\bf Definition 1}.
\hspace{1mm}
The random matrix $\boldX$ ($p \times n$) is said to have a matrix variate normal distribution with mean matrix $\boldM$ ($p \times n$) and covariance matrix $\boldPsi \otimes \boldSigma$ where $\boldPsi$ $(p \times p)$ and $\boldSigma$ $(n \times n)$ are positive definite matrices if $\vecm{(\boldX')} \sim \Normal(\vecm{(\boldM')}, \boldPsi \otimes \boldSigma)$ and we denote $\boldX \sim \Normal_{p,n}(\boldM, \boldPsi \otimes \boldSigma)$.
\subsection{Modeling the leverage effect}
We extend our proposed model to incorporate the leverage effect as follows. The joint distribution of $(\boldy_t, \boldh_{t+1})$ is given by
\begin{align}
\begin{pmatrix}
\boldy_t \\
\boldh_{t+1} 
\end{pmatrix}
\sim \Normal
\left(
\begin{pmatrix}
\boldm_t \\
\boldmu + \boldPhi (\boldh_{t} - \boldmu)
\end{pmatrix}
,
\begin{pmatrix}
\boldV_t^{1/2}\boldR_t\boldV_t^{1/2}  & \boldV_t^{1/2} \boldR_t^{1/2} \boldLambda' \\
\boldLambda\boldR_t^{1/2'} \boldV_t^{1/2} & \boldPsi + \boldLambda\boldLambda'
\end{pmatrix}
\right)
.
\end{align}
The marginal distributions of $\bm{y}_t$ and $\bm{h}_{t+1}$ given $\bm{h}_t$ are the same as before with  $\mathbf{\Omega}=\mathbf{\Psi}+\boldLambda\boldLambda'$, but we note that
\begin{eqnarray*}
\bm{h}_{t+1}|\bm{y}_t, \bm{h}_t, \bm{\theta} \sim 
N\left(\boldmu + \boldPhi(\boldh_t - \boldmu) + \boldLambda \boldR_t^{-1/2}\boldV_t^{-1/2} (\boldy_t - \boldm_t),  \mathbf{\Psi}\right).
\end{eqnarray*}
If $\mathbf{\Lambda}=\mathbf{O}$, it reduces to the model without leverage effect. The matrix $\mathbf{\Lambda}$ is the coefficient of the leverage for $\bm{z}_t = \mathbf{R}_t^{-1/2}\mathbf{V}_t^{-1/2}(\bm{y}_t-\bm{m}_t)$.
 We assume that the prior distribution of $\mathbf{\Lambda}$ given $\mathbf{\Psi}$ is $\Normal_{p,p}(\boldM_0, \boldPsi \otimes \boldGamma_0)$. That is, 
$\mathbf{\Lambda} |\mathbf{\Psi} \sim \Normal_{p,p}(\boldM_0, \boldPsi \otimes \boldGamma_0).$
\\

\noindent
{\it Remark} 3. There are several ways to choose $\boldR_t^{1/2}$. For example, we can use the spectral decomposition of the correlation matrix $\boldR_t=\boldP_t\boldQ_t\boldP_t'$ and $\boldR_t^{1/2} = \boldP_t \boldQ_t^{1/2}$ where the $i$-th diagonal element of the diagonal matrix $\boldQ_t$ is the $i$-th largest eigenvalue of $\boldR_t$ and the $i$-th column of $\boldP_t$ is the corresponding $i$-th eigenvector (and we set the first elements of the eigenvectors to be positive for the identification purpose).  Thus the $i$-th element of $\bm{z}_t$ can be interpreted as the $i$-th market factor among $p$ asset returns. Alternatively, Cholesky decomposition, $\boldR_t = \boldR_t^{1/2}\boldR_t^{1/2'}$,  can be used so that $\boldR_t^{1/2}$ is a lower triangular matrix where all the diagonal elements are equal to one, but we note that it is affected by the ordering of the asset returns.

\subsubsection{Generation of $\mathbf{\Lambda}$}
The conditional posterior distribution of $\mathbf{\Lambda}$ is derived in the following Proposition and we generate $\vecm(\boldLambda) \vert \cdot \sim \Normal(\vecm(\boldM_1'), \boldPsi \otimes \boldGamma_1)$.\\

\noindent
{\bf Proposition 2}.
\hspace{1mm}
Suppose that the prior distribution of $\mathbf{\Lambda}$ given $\mathbf{\Psi}$ is $\Normal_{p,p}(\boldM_0, \boldPsi \otimes \boldGamma_0)$. Then the conditional posterior distribution of $\mathbf{\Lambda}$ given other parameters and latent variables is $\Normal_{p,p}(\boldM_1, \boldPsi \otimes \boldGamma_1)$
where
\begin{eqnarray}
\boldM_1 & = & \left(\boldA + \boldGamma_0^{-1}\right)^{-1} \left(\boldB + \boldGamma_0^{-1} \boldM_0\right), \quad \mathbf{\Gamma}_1 =\left(\boldA + \boldGamma_0^{-1}\right)^{-1},\\
\label{eq:AB}
\boldA & = & \sum_{t=1}^{T-1}\bm{z}_t\bm{z}_t',
\quad \boldB =\sum_{t=1}^{T-1}\bm{z}_t\bm{\eta}_t',
\end{eqnarray}
and
$\bm{z}_t = \mathbf{R}_t^{-1/2}\mathbf{V}_t^{-1/2}(\bm{y}_t-\bm{m}_t)$ and
$\bm{\eta}_t=\bm{h}_{t+1}-\bm{\mu}-\mathbf{\Phi}(\bm{h}_t-\bm{\mu})$.
\\
\noindent
{\bf Proof:} See Appendix \ref{sec:proof2}.
\\
\noindent
For the generations of other parameters and latent variables, see Appendix \ref{sec:mcmc_w_leverage}
.
\subsection{Parsimonious specification of the leverage effect}
This subsection proposes the parsimonious specification for $\boldLambda= [\boldlambda_1, \cdots, \bm{\lambda}_p]$, in order to reduce the number of leverage parameters from $p^2$ to $pq$ $(q\ll p)$ by setting $\boldLambda = [\boldlambda_1, \cdots, \bm{\lambda}_q,\boldzero,\ldots,\boldzero]$ since we do not have additional measurement equations for the leverage effect. Using the spectral decomposition to compute $\mathbf{R}_t^{1/2}$, we can interpret that the $i$-th column corresponds to the $i$-th market factor among asset returns $(i=1,\ldots,q)$. The number of factors, $q$, is expected to be small, e.g., $q=1$ or $q=2$. 
\subsubsection{Generation of $\mathbf{\Lambda}= [\boldlambda_1, \cdots, \bm{\lambda}_q,\boldzero,\ldots,\boldzero]$}
The following proposition and the corollary shows the conditional posterior distribution of the parameters for the leverage effect under parsimonious specifications.
\\

\noindent
{\bf Proposition 3}.
Let $\boldLambda = [\boldlambda_1, \cdots, \bm{\lambda}_q,\boldzero,\ldots,\boldzero]$ and $\bm{\lambda}=(\bm{\lambda}_1',\ldots,\bm{\lambda}_q')'$. If the prior distribution of $\bm{\lambda}$ is assumed to be normal, $\boldlambda \sim \Normal(\boldm_{0}, \boldGamma_{0})$, then the conditional posterior distribution of $\boldlambda$ is $\boldlambda \vert \cdot \sim \Normal(\boldm_{1},\boldGamma_{1})$ where
\begin{align*}
&\boldm_{1} =  
	\boldGamma_{1}
	\left\{
	\boldGamma_{0}^{-1} \bm{m}_{0}+(\mathbf{I}_q\otimes\boldPsi^{-1}\mathbf{B}') \text{vec}\left(\{\bolde_1,\ldots,\bolde_q\}\right)
	\right\},
	\quad
\boldGamma_{1}= \left(\boldGamma_{0}^{-1}+\mathbf{A}_{1:q,1:q}\otimes\ \boldPsi^{-1}\right)^{-1}, 
\end{align*}
$\mathbf{A},\mathbf{B}$ are defined in (\ref{eq:AB}), 
$\boldA_{1:q,1:q}$ denotes the first $q$ rows and the $q$ columns of $\boldA$,
$\text{vec}(\mathbf{X})\equiv (\bm{x}_1', \ldots, \bm{x}_m')'$ denotes a vectorization of the matrix $\mathbf{X}=\{\bm{x}_1, \ldots, \bm{x}_m\}$, and $\otimes$ denotes Kronecker product.
\\
\noindent
{\bf Proof:} See Appendix \ref{sec:proof3}.
\\

\noindent
{\bf Corollary 1}.
Let $q=1$ and $\boldLambda = [\boldlambda, \boldzero,\ldots,\boldzero]$. If the prior distribution of $\bm{\lambda}$ is assumed to be normal, $\boldlambda \sim \Normal(\boldm_{0}, \boldGamma_{0})$, then the conditional posterior distribution of $\boldlambda$ is $\boldlambda \vert \cdot \sim \Normal(\boldm_{1},\boldGamma_{1})$ where
\begin{eqnarray}
\boldm_{1}  & = &   
	\boldGamma_{1}
	\left\{
	\boldGamma_{0}^{-1} \bm{m}_{0}+ \boldPsi^{-1}\bm{b}
	\right\}, 
	\quad \boldGamma_{1}= \left(\boldGamma_{0}^{-1}+a\times\boldPsi^{-1} \right)^{-1},
\\
&&
a =\sum_{t=1}^{T-1}z_{1t}^2, \quad 
\bm{b} = \sum_{t=1}^{T-1}z_{1t}
	\left\{
	\bm{h}_{t+1}-\bm{\mu}-\mathbf{\Phi}(\bm{h}_t-\bm{\mu})
	\right\},
\nonumber
\end{eqnarray}
and $z_{1t}$ is the first element of $\bm{z}_t = \mathbf{R}_t^{-1/2}\mathbf{V}_t^{-1/2}(\bm{y}_t-\bm{m}_t)$.
\subsubsection{Generation of $\mathbf{\Psi}$}
See Appendix \ref{sec:mcmc_w_leverage}.

\section{Empirical studies}
This section applies our proposed model to the daily returns of nine U.S. stocks ($p=9$) with realized volatilities and pairwise realized correlations. The nine series of stock returns are JP Morgan (JPM), International Business Machine (IBM), Microsoft (MSFT), Exxon Mobil (XOM), Alcoa (AA), American Express (AXP), Du Pont (DD), General Electric (GE), and Coca Cola (KO).
The sample period is from February 1, 2001 to December 31, 2009, and the number of observation is $T=2242$. The daily returns for the $i$-th stocks are defined as $y_{it} = 100 \times (\log p_{it} - \log p_{i, t-1})$, where $p_{it}$ is the closing price of the $i$-th asset at time $t$. 
\begin{figure}[H]
\centering 
\hspace{-10mm}
\includegraphics[width = 0.45\linewidth, angle = 270]{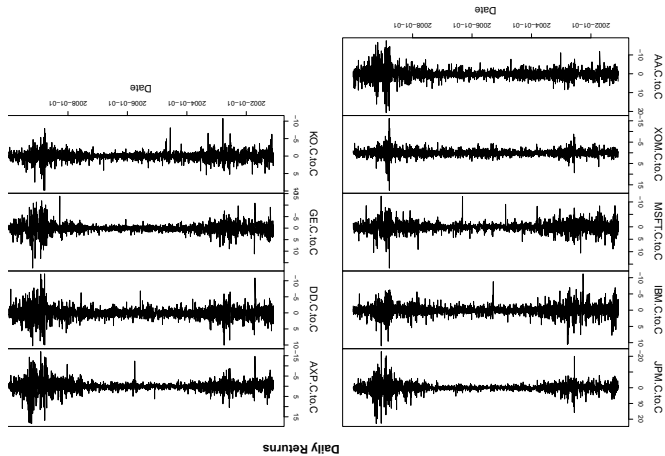}\caption{Time series plots of nine US stock (close-to-close) returns.}
\label{figure:realreturn}
\end{figure}
\noindent
Time series plots of $y_{it}$ are shown in Figure \ref{figure:realreturn}, and show that there is a very high volatility period in 2008 (the financial crisis when Lehman Brothers filed for Chapter 11 bankruptcy protection). Additionally, there are other relatively high volatility periods in 2001 (the dot-com bubble and the September 11 attacks) and in 2002 (the market turmoil during which Worldcom filed for Chapter 11 bankruptcy protection).
The realized volatilities and pairwise realized correlations are computed from the realized covariance matrices for these assets which can be downloaded from the Oxford Man Institute website (see, Section 5 of \cite{noureldin2012multivariate} for details). 
The prior distributions are assumed to be vague and flat in order to reflect the fact that we have little information with regard to the parameters:
\begin{align*}
&\mu_i \sim \Normal(0, 10^4),  \quad \xi_i \sim \Normal(0, 10^4),
\quad \delta_{ij} \sim \Normal(0, 10^4),  \quad \lambda_i \sim \Normal(0, 10^4),
\\
&\frac{1+\phi_i}{2} \sim \text{Beta}(1, 1),\quad \sigma_{u,i}^2 \sim \IG(10^{-5}/2, 10^{-5}/2), \quad \sigma_{v,ij}^2 \sim \IG(10^{-5}/2, 10^{-5}/2),
\\
&\sigma_{\zeta,ij}^2 \sim \IG(10^{-6}/2, 10^{-6}/2), \quad \sigma_{m,i}^2 \sim \IG(10^{-5}/2, 10^{-5}/2), \quad 
\quad \mathbf{\Psi} \sim \IW(10, \boldI_{9}),
\end{align*}
for $i=1,\ldots,p$, $j= 1,\ldots,i-1$.
The proposed model is estimated and we use the parsimonious specification of the leverage effect with the number of factors $q=1$.

\subsection{Estimation results}
We run 12,000 MCMC iterations and the first 2,000 iterations are discarded as the burn-in period.
Table \ref{table:empirical_1} shows the posterior means, 95\% credible intervals and inefficiency factors\footnote{The inefficiency factor is defined as $1+2\sum_{g=1}^\infty \rho(g)$,
where $\rho(g)$ is the sample autocorrelation at lag $g$.  This is interpreted as the ratio of the numerical variance of the posterior mean from the chain to the variance of the posterior mean from hypothetical uncorrelated draws. The smaller the inefficiency factor becomes, the closer the MCMC sampling is to the uncorrelated sampling.} for $\boldmu, \boldxi, \boldphi,\boldsigma_u,\boldsigma_m$ and $\bm{\lambda}$. The inefficiency factors are relatively small (less than 130 ) in the multivariate stochastic volatility models and our algorithm works well. The posterior means and posterior standard deviations for $\bm{\delta}$, $\bm{\sigma}_v$ and $\bm{\sigma}_{\zeta}$
are shown in Tables \ref{table:empirical_delta},  \ref{table:empirical_sigv} and \ref{table:empirical_sigzeta}, respectively.
\\
\noindent
{\it Mean processes and volatilities}. The posterior means of $\sigma_{m,i}$ are around $0.067\sim 0.098$, which reflects that the magnitude of the mean process, $\bm{m}_t$, is much smaller than that of the stochastic volatility component, $\mathbf{V}_t^{1/2}\bm{\epsilon}_t$, as we expected. The unconditional means of the log volatilities, $\mu_i$,  are estimated to be from $0.203$ to $1.582$ and the posterior mean of $\mu_5$ (corresponding to Alcoa) is much larger than those of others. The stock returns of Alcoa are found to be the most volatile among others, while those of Coca Cola are the least volatile. 

\begin{figure}[H]
\centering 
\includegraphics[width = 0.725\linewidth, angle = 0]{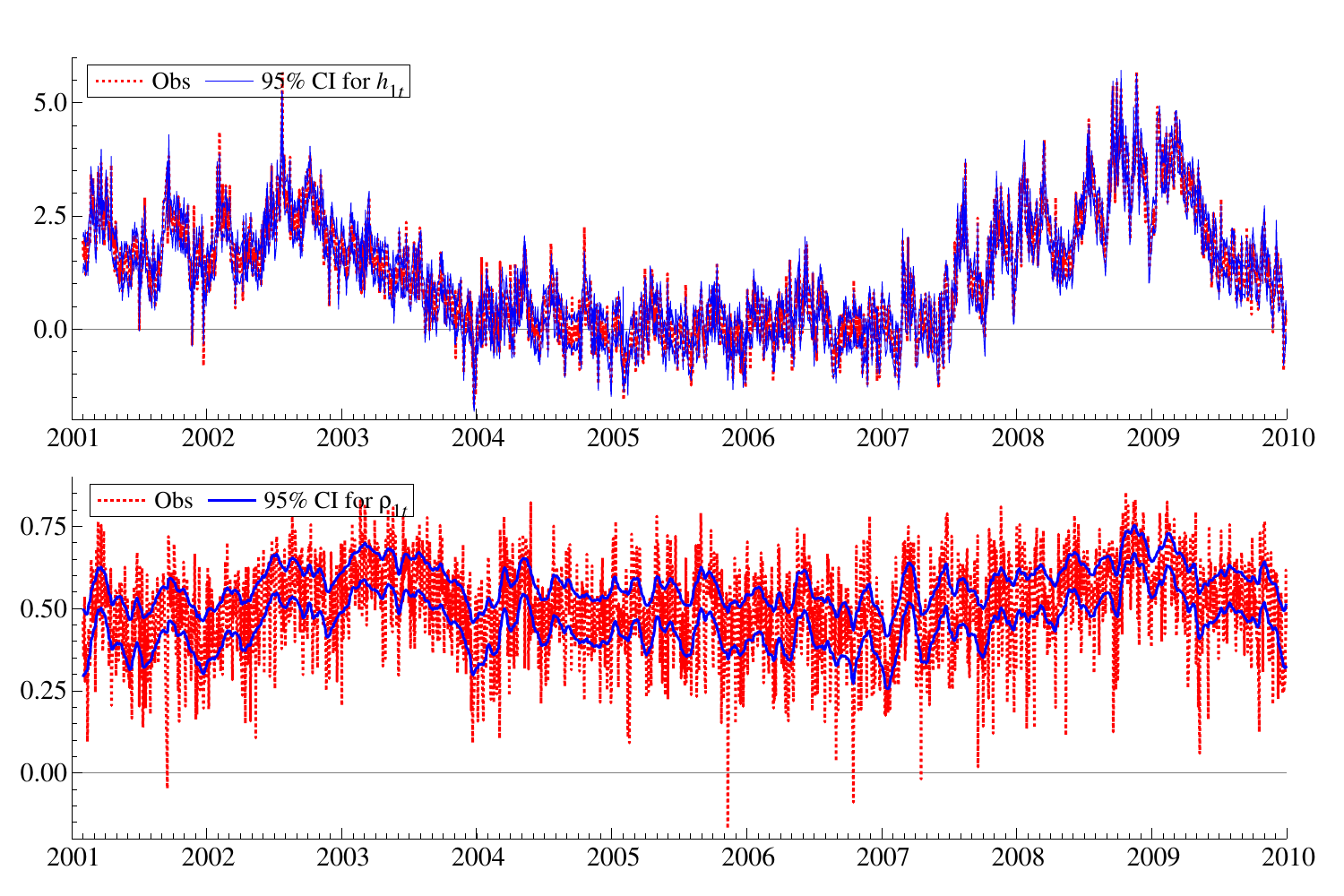}
\vspace{-4mm}
\caption{Top: 95\% credible intervals (solid) of $h_{1t}$,  and $x_{1t} - \xi_1$ (dotted).
Bottom: 95\% credible intervals (solid) of $\rho_{21,t}$ and $\{\exp(w_{21,t}-\delta_{21})-1\}/\{\exp(w_{21,t}-\delta_{21})+1\}$ (dotted).}
\label{figure:hx_el21}
\end{figure}

%
%
%
%
%
%
\begin{table}[H]
\footnotesize
\caption{Posterior means, 95\% credible intervals and inefficiency} factors for $\bm{\mu}$, $\bm{\xi}$, $\bm{\phi}$, $\bm{\sigma}_u$, $\bm{\sigma}_m$ and $\bm{\lambda}$.
Spectral decomposition is used to compute $\mathbf{R}_t^{-1/2}$.
\label{table:empirical_1}
\vspace{5mm} \\
\centering
\scalebox{1.0}{
\begin{tabular}{rrcrrrcr}
\hline\hline
\multicolumn{1}{l}{Par.}&\multicolumn{1}{c}{Mean}&\multicolumn{1}{c}{95\% interval}&\multicolumn{1}{c}{IF} & 
\multicolumn{1}{l}{Par.}&\multicolumn{1}{c}{Mean}&\multicolumn{1}{c}{95\% interval}&\multicolumn{1}{c}{IF}
\tabularnewline
\hline

$\mu_{1}$&$1.221$&[1.020,1.430] &$ 8$&$\sigma_{u,1}$&$0.285$&[0.270,0.300]&$40$\tabularnewline
$\mu_{2}$&$0.644$&[0.495,0.794] &$16$&$\sigma_{u,2}$&$0.286$&[0.271,0.303]&$57$\tabularnewline
$\mu_{3}$&$0.914$&[0.759,1.070] &$ 9$&$\sigma_{u,3}$&$0.291$&[0.278,0.304]&$14$\tabularnewline
$\mu_{4}$&$0.684$&[0.541,0.827] &$11$&$\sigma_{u,4}$&$0.274$&[0.261,0.287]&$20$\tabularnewline
$\mu_{5}$&$1.582$&[1.430,1.730] &$ 8$&$\sigma_{u,5}$&$0.318$&[0.304,0.332]&$17$\tabularnewline
$\mu_{6}$&$1.136$&[0.918,1.360] &$ 9$&$\sigma_{u,6}$&$0.307$&[0.293,0.321]&$20$\tabularnewline
$\mu_{7}$&$0.873$&[0.726,1.020] &$12$&$\sigma_{u,7}$&$0.291$&[0.278,0.304]&$20$\tabularnewline
$\mu_{8}$&$0.861$&[0.670,1.050] &$10$&$\sigma_{u,8}$&$0.303$&[0.289,0.317]&$18$\tabularnewline
$\mu_{9}$&$0.203$&[0.055,0.352] &$15$&$\sigma_{u,9}$&$0.294$&[0.281,0.308]&$18$\vspace{2mm}\tabularnewline

$\xi_{1}$&$-0.520$&[-0.582,-0.470] &$102$&$\sigma_{m,1}$&$0.087$&[0.067,0.108] &$118$\tabularnewline
$\xi_{2}$&$-0.554$&[-0.610,-0.495] &$101$&$\sigma_{m,2}$&$0.070$&[0.053,0.089] &$126$\tabularnewline
$\xi_{3}$&$-0.549$&[-0.594,-0.501] &$ 90$&$\sigma_{m,3}$&$0.076$&[0.053,0.106] &$129$\tabularnewline
$\xi_{4}$&$-0.442$&[-0.487,-0.394] &$ 88$&$\sigma_{m,4}$&$0.087$&[0.067,0.106] &$117$\tabularnewline
$\xi_{5}$&$-0.537$&[-0.582,-0.494] &$ 78$&$\sigma_{m,5}$&$0.098$&[0.073,0.135] &$129$\tabularnewline
$\xi_{6}$&$-0.586$&[-0.651,-0.533] &$109$&$\sigma_{m,6}$&$0.088$&[0.071,0.118] &$119$\tabularnewline
$\xi_{7}$&$-0.428$&[-0.480,-0.376] &$102$&$\sigma_{m,7}$&$0.087$&[0.063,0.111] &$123$\tabularnewline
$\xi_{8}$&$-0.535$&[-0.589,-0.474] &$100$&$\sigma_{m,8}$&$0.078$&[0.060,0.108] &$125$\tabularnewline
$\xi_{9}$&$-0.322$&[-0.376,-0.263] &$ 93$&$\sigma_{m,9}$&$0.067$&[0.048,0.088] &$124$\vspace{2mm}\tabularnewline

$\phi_{1}$&$0.914$&[0.904,0.924] &$19$&$\lambda_{1}$&$-0.0626$&[-0.0852,-0.0403] &$7$\tabularnewline
$\phi_{2}$&$0.888$&[0.874,0.901] &$22$&$\lambda_{2}$&$-0.0541$&[-0.0757,-0.0330] &$7$\tabularnewline
$\phi_{3}$&$0.900$&[0.887,0.913] &$19$&$\lambda_{3}$&$-0.0430$&[-0.0638,-0.0216] &$7$\tabularnewline
$\phi_{4}$&$0.890$&[0.876,0.904] &$21$&$\lambda_{4}$&$-0.0518$&[-0.0722,-0.0311] &$6$\tabularnewline
$\phi_{5}$&$0.907$&[0.895,0.920] &$20$&$\lambda_{5}$&$-0.0424$&[-0.0625,-0.0219] &$7$\tabularnewline
$\phi_{6}$&$0.926$&[0.916,0.935] &$25$&$\lambda_{6}$&$-0.0518$&[-0.0735,-0.0303] &$6$\tabularnewline
$\phi_{7}$&$0.899$&[0.886,0.911] &$22$&$\lambda_{7}$&$-0.0536$&[-0.0736,-0.0331] &$9$\tabularnewline
$\phi_{8}$&$0.908$&[0.897,0.920] &$21$&$\lambda_{8}$&$-0.0538$&[-0.0767,-0.0308] &$8$\tabularnewline
$\phi_{9}$&$0.903$&[0.889,0.916] &$21$&$\lambda_{9}$&$-0.0436$&[-0.0637,-0.0235] &$10$\tabularnewline
\hline
\end{tabular}
}
\normalsize
\end{table}
\noindent
Since all posterior means of the autoregressive coefficients, $\phi_i$, are approximately 0.9, the log volatilities are found to have high persistence. The elements of $\mathbf{\Psi}$ (the conditional covariance matrix of $\bm{h}_{t+1}$ given $\bm{y}_t$) are all approximately 0.1 and the probability that $\psi_{ij}$ is positive is greater than 0.975 for all $i$s and $j$s. The log volatilities, $h_{i,t+1}$, are positively correlated with each other given $\bm{y}_t$.
Figure \ref{figure:hx_el21} shows the 95\% credible intervals for $h_{1t}$ with $x_{1t}-\xi_1$ where $\xi_1$ is the estimated posterior mean of the first bias correction term.
The figures for $h_{it}$ $(i=2,\ldots,9)$ are similar and hence are omitted.  
The estimated 95 \% credible intervals have smaller fluctuation than those of the bias-adjusted realized measures. These estimates succeeded at automatically extracting the mean trends of the volatilities and adjusting the measurement errors. Overall, the 95\% credible intervals captures the traceplot of the (bias-corrected) realized volatilities, suggesting that our proposed model is successful at describing the dynamics of the latent log volatilities.
\vspace{-5mm}
\noindent
%
%
%
%
\begin{table}[H]
\vspace{-8mm}
\footnotesize
\newcolumntype{l}{>{\centering}p{4em}}
\caption{Posterior means (posterior standard deviations) of $\bm{\delta}$. \label{table:empirical_delta}} 
\vspace{2mm} 
\centering
\scalebox{0.835}{
\begin{tabular}{lllllllll}
\hline\hline
\multicolumn{1}{l}{$\delta_{ij}$}&\multicolumn{1}{c}{$j = 1$}&\multicolumn{1}{c}{$j = 2$}&\multicolumn{1}{c}{$j = 3$}&\multicolumn{1}{c}{$j = 4$}&\multicolumn{1}{c}{$j = 5$}&\multicolumn{1}{c}{$j = 6$}&\multicolumn{1}{c}{$j = 7$}&\multicolumn{1}{c}{$j = 8$}\tabularnewline
\hline
$i = 2$&{\bf-0.313}\newline(0.029)&&&&&&&\tabularnewline
$i = 3$&{\bf-0.265}\newline(0.031)&{\bf-0.333}\newline(0.053)&&&&&&\tabularnewline
$i = 4$&{\bf-0.301}\newline(0.048)&{\bf-0.180}\newline(0.035)&{\bf-0.149}\newline(0.065)&&&&&\tabularnewline
$i = 5$&{\bf-0.410}\newline(0.065)&{\bf-0.227}\newline(0.027)&{\bf-0.246}\newline(0.038)&{\bf-0.287}\newline(0.043)&&&&\tabularnewline
$i = 6$&{\bf-0.629}\newline(0.040)&{\bf-0.265}\newline(0.032)&{\bf-0.295}\newline(0.048)&{\bf-0.353}\newline(0.045)&{\bf-0.391}\newline(0.036)&&&\tabularnewline
$i = 7$&{\bf-0.472}\newline(0.047)&{\bf-0.276}\newline(0.029)&{\bf-0.219}\newline(0.030)&{\bf-0.312}\newline(0.025)&{\bf-0.532}\newline(0.035)&{\bf-0.477}\newline(0.032)&&\tabularnewline
$i = 8$&{\bf-0.526}\newline(0.053)&{\bf-0.370}\newline(0.041)&{\bf-0.319}\newline(0.065)&{\bf-0.340}\newline(0.037)&{\bf-0.440}\newline(0.029)&{\bf-0.563}\newline(0.048)&{\bf-0.478}\newline(0.033)&\tabularnewline
$i = 9$&{\bf-0.194}\newline(0.043)&{\bf-0.081}\newline(0.033)&{\bf-0.098}\newline(0.026)&{\bf-0.249}\newline(0.050)&{\bf-0.150}\newline(0.029)&{\bf-0.246}\newline(0.030)&{\bf-0.158}\newline(0.026)&{\bf-0.172}\newline(0.057)\tabularnewline
\hline
\end{tabular}
}
\vspace{1mm}\\
\footnotesize
*Bold figures indicate that the 95\% credible interval does not include zero.
\normalsize
\end{table}
%
%
%
%
%
\begin{table}[H]
\vspace{-3mm}
\footnotesize
\newcolumntype{l}{>{\centering}p{4em}}
\caption{Posterior means (posterior standard deviations) of $\bm{\sigma}_{v}$. \label{table:empirical_sigv}} \vspace{2mm} 
\centering
\scalebox{0.835}{
\begin{tabular}{lllllllll}
\hline\hline
\multicolumn{1}{l}{$\sigma_{v,ij}$}&\multicolumn{1}{c}{$j = 1$}&\multicolumn{1}{c}{$j = 2$}&\multicolumn{1}{c}{$j = 3$}&\multicolumn{1}{c}{$j = 4$}&\multicolumn{1}{c}{$j = 5$}&\multicolumn{1}{c}{$j = 6$}&\multicolumn{1}{c}{$j = 7$}&\multicolumn{1}{c}{$j = 8$}\tabularnewline
\hline
$i = 2$&0.329\newline(0.006)&&&&&&&\tabularnewline
$i = 3$&0.325\newline(0.005)&0.325\newline(0.005)&&&&&&\tabularnewline
$i = 4$&0.329\newline(0.005)&0.329\newline(0.006)&0.310\newline(0.005)&&&&&\tabularnewline
$i = 5$&0.329\newline(0.005)&0.322\newline(0.005)&0.314\newline(0.005)&0.328\newline(0.005)&&&&\tabularnewline
$i = 6$&0.350\newline(0.006)&0.331\newline(0.006)&0.315\newline(0.005)&0.319\newline(0.006)&0.332\newline(0.006)&&&\tabularnewline
$i = 7$&0.338\newline(0.006)&0.338\newline(0.006)&0.319\newline(0.005)&0.334\newline(0.006)&0.344\newline(0.006)&0.338\newline(0.006)&&\tabularnewline
$i = 8$&0.334\newline(0.006)&0.320\newline(0.006)&0.304\newline(0.005)&0.322\newline(0.006)&0.317\newline(0.005)&0.332\newline(0.006)&0.326\newline(0.005)&\tabularnewline
$i = 9$&0.314\newline(0.005)&0.329\newline(0.006)&0.307\newline(0.005)&0.317\newline(0.006)&0.320\newline(0.005)&0.321\newline(0.005)&0.337\newline(0.006)&0.328\newline(0.006)\tabularnewline
\hline
\end{tabular}
}
\normalsize
\end{table}
\vspace{-5mm}
%
%
%
%

\begin{table}[H]
\footnotesize
\newcolumntype{l}{>{\centering}p{4em}}
\caption{Posterior means (posterior standard deviations) of $\bm{\sigma}_{\zeta}$.\label{table:empirical_sigzeta}} 
\vspace{2mm} 
\centering
\scalebox{0.835}{
\begin{tabular}{lllllllll}
\hline\hline
\multicolumn{1}{l}{$\sigma_{\zeta,ij}$}&\multicolumn{1}{c}{$j = 1$}&\multicolumn{1}{c}{$j = 2$}&\multicolumn{1}{c}{$j = 3$}&\multicolumn{1}{c}{$j = 4$}&\multicolumn{1}{c}{$j = 5$}&\multicolumn{1}{c}{$j = 6$}&\multicolumn{1}{c}{$j = 7$}&\multicolumn{1}{c}{$j = 8$}\tabularnewline
\hline
$i = 2$&0.047\newline(0.005)&&&&&&&\tabularnewline
$i = 3$&0.039\newline(0.004)&0.043\newline(0.004)&&&&&&\tabularnewline
$i = 4$&0.061\newline(0.004)&0.062\newline(0.006)&0.061\newline(0.005)&&&&&\tabularnewline
$i = 5$&0.047\newline(0.004)&0.044\newline(0.004)&0.042\newline(0.004)&0.047\newline(0.004)&&&&\tabularnewline
$i = 6$&0.058\newline(0.005)&0.046\newline(0.005)&0.044\newline(0.005)&0.069\newline(0.006)&0.047\newline(0.004)&&&\tabularnewline
$i = 7$&0.041\newline(0.004)&0.044\newline(0.005)&0.041\newline(0.005)&0.061\newline(0.005)&0.053\newline(0.005)&0.049\newline(0.005)&&\tabularnewline
$i = 8$&0.048\newline(0.005)&0.052\newline(0.005)&0.055\newline(0.005)&0.071\newline(0.005)&0.049\newline(0.004)&0.054\newline(0.006)&0.049\newline(0.005)&\tabularnewline
$i = 9$&0.049\newline(0.005)&0.042\newline(0.005)&0.040\newline(0.004)&0.064\newline(0.006)&0.043\newline(0.004)&0.045\newline(0.005)&0.044\newline(0.005)&0.048\newline(0.004)\tabularnewline
\hline
\end{tabular}
}
\normalsize
\end{table}
\noindent
{\it Biases in realized volatilities and correlations}.
The bias correction terms, $\xi_i$, of the realized volatilities are estimated to be negative, thereby indicating that the realized volatilities have downward biases and underestimate the  volatilities by ignoring the overnight nontrading hours. Since the realized volatilities tend to overestimate the volatilities due to the microstructure noises, the effect of nontrading hours seems to dominate in the direction of the biases. We also note that the magnitudes of the biases depend on the series of stock returns. 
Table \ref{table:empirical_delta} shows the estimation result of the bias term $\bolddelta$ of the correlation coefficients. All $\delta_{ij}$ are estimated to be negative, and the posterior probability that $\delta_{ij}$ is negative is greater than 0.975. This implies that the realized correlations underestimate the latent correlations, thereby suggesting the existence of the Epps effect. 
\vspace{2mm}
\\
\noindent
{\it Dynamic correlations}. The posterior means of the standard deviations of the disturbance terms in the state equations corresponding to the dynamic correlations, $\sigma_{\zeta,ij}$, are shown in Table \ref{table:empirical_sigzeta}. They are $0.039\sim 0.071$ and are much smaller than the posterior means of the standard deviations for the measurement errors of the realized measures, $\sigma_{u,i}$ and $\sigma_{v,ij}$ (as shown in Tables \ref{table:empirical_1} and \ref{table:empirical_sigv}) which are found to be similar for all $i$s and $j$s at approximately 0.30. 
Figure \ref{figure:hx_el21}
shows the time series plots of the 95\% credible intervals of the selected dynamic correlations, $\rho_{21,t}$ with $\{\exp(w_{21,t}-\delta_{21})-1\}/\{\exp(w_{21,t}-\delta_{21})+1\}$ where $\delta_{21}$ is the estimated posterior mean. The figures for the other $\rho_{ij,t}$ are similar and hence are omitted. Again, the estimated 95\% credible intervals  of $\rho_{ij,t}$ have much smaller fluctuation those of bias-adjusted realized measures, $\{\exp(x_{ij,t}-\delta_{ij})-1\}/\{\exp(x_{ij,t}-\delta_{ij})+1\}$. These intervals seem to extract the mean trends of the bias-adjusted realized measures that have relatively large noises in the measurement equation. The correlations between the asset returns are found to be time-varying in the sample period, and they seem to increase after the financial crisis in 2008.
This result corresponds to our intuition that each asset return has a larger positive correlation with others when the market faces stress, rather than when it is in a usual period. 
\vspace{2mm}
\\
\noindent
{\it Leverage effect and the selection of the number of factors $q$}. 
The parameters for the leverage effect, $\lambda_i$, are estimated to be negative in Table \ref{table:empirical_1} and the posterior probability that $\lambda_i$ is negative is greater than 0.975 for all $i$s. This implies the existence of the leverage effect. 
Table \ref{table:empirical_leverage} also shows the estimation results for the correlation between the first element of $\bm{z}_t=\mathbf{V}_t^{-1/2}\mathbf{R}_t^{-1/2}\bm{y}_t$ and $h_{i,t+1}$, {\it i.e.},  $\rho_i^*=Corr(z_{1t}, h_{i,t+1} )=\lambda_{ii}/\sqrt{\lambda_{ii}^2+\psi_{ii}}$ for $i=1,\ldots,9$. The posterior means of $\rho_i^*$ are estimated to be negative ranging from $-0.22$ to $-0.15$. If we regard $z_{1t}$ as the market factor, a decrease in the market return ($z_{1t}$) is followed by an increase in the log volatility ($h_{i,t+1}$), which implies the existence of the leverage effect. The estimation results for $\mathbf{\Psi}$ are omitted in order to save space where the posterior probability that $\psi_{ij}>0$ is found to be greater than 0.975 for all $i$s and $j$s.

To investigate whether the number of factors is $q=1$, we also fit the proposed model using $q=2$ and we set  $\mathbf{\Lambda}=[\bm{\lambda}_1,\bm{\lambda}_2,\bm{0},\ldots,\bm{0}]$ for the leverage effect.  Table \ref{table:empirical_leverage2} shows the posterior means, the 95\% credible intervals and inefficiency factors for $\rho_{1i}^*=Corr(z_{1t}, h_{i,t+1})$ and $\rho_{2i}^*=Corr(z_{2t}, h_{i,t+1})$ for $i=1,\ldots,9$ where $\bm{z}_t=\mathbf{R}_t^{-1/2}\mathbf{V}_t^{-1/2}(\bm{y}_t-\bm{m}_t)$. The estimation results for $\rho_{1i}^*$  are almost the same as those for $\rho_i^*$ in Table \ref{table:empirical_leverage}. Conversely, the posterior means of $\rho_{2i}^*$ are close to zeros, and the 95\% credible intervals include zero. This suggests that one factor ($q=1$) is enough to describe the leverage effect for our dataset.

%
%
%
%
\begin{table}[H]
\vspace{-2mm}
\footnotesize
\centering
\caption{Posterior means, 95\% credible intervals and inefficiency factors for  $\rho_i^*=Corr(z_{1t}, h_{i,t+1})$ where $\bm{z}_t=\mathbf{R}_t^{-1/2}\mathbf{V}_t^{-1/2}(\bm{y}_t-\bm{m}_t)$ and $q=1$. Spectral decomposition is used.}
\vspace{2mm} 
\centering
\label{table:empirical_leverage}
\scalebox{1.0}{
\begin{tabular}{lllr}
\hline\hline
\multicolumn{1}{l}{Par.}&\multicolumn{1}{c}{Mean}&\multicolumn{1}{c}{95\% interval}&\multicolumn{1}{c}{IF}\tabularnewline
\hline
$\rho_{1}^*$&$-0.199$&[-0.306,-0.113]&$ 8$\tabularnewline
$\rho_{2}^*$&$-0.187$&[-0.302,-0.100]&$ 8$\tabularnewline
$\rho_{3}^*$&$-0.153$&[-0.261,-0.067]&$ 7$\tabularnewline
$\rho_{4}^*$&$-0.199$&[-0.327,-0.101]&$ 7$\tabularnewline
$\rho_{5}^*$&$-0.179$&[-0.322,-0.076]&$ 8$\tabularnewline
$\rho_{6}^*$&$-0.177$&[-0.288,-0.089]&$ 8$\tabularnewline
$\rho_{7}^*$&$-0.224$&[-0.374,-0.114]&$10$\tabularnewline
$\rho_{8}^*$&$-0.171$&[-0.278,-0.087]&$ 8$\tabularnewline
$\rho_{9}^*$&$-0.187$&[-0.334,-0.083]&$11$\tabularnewline
\hline
\end{tabular}
}
\normalsize
\end{table}
\vspace{-4mm}
\begin{table}[H]
\footnotesize
\centering
\caption{Posterior means, 95\% credible intervals and inefficiency factors for $\rho_{1i}^*=Corr(z_{1t}, h_{i,t+1})$ and $\rho_{2i}^*=Corr(z_{2t}, h_{i,t+1})$ where $q=2$. Spectral decomposition is used.}
\vspace{2mm} 
\label{table:empirical_leverage2}
\scalebox{1.0}{
\begin{tabular}{lrcrlrcl}
\hline\hline
\multicolumn{1}{l}{Par.}&\multicolumn{1}{c}{Mean}&\multicolumn{1}{c}{95\% interval}&\multicolumn{1}{c}{IF} & 
\multicolumn{1}{l}{Par.}&\multicolumn{1}{c}{Mean}&\multicolumn{1}{c}{95\% interval}&\multicolumn{1}{c}{IF}\tabularnewline
\hline
$\rho_{11}^*$&$-0.195$&[-0.305,-0.107] &$ 6$&$\rho_{21}^*$&$-0.009$&[-0.103,0.071] &$35$\tabularnewline
$\rho_{12}^*$&$-0.183$&[-0.305,-0.091] &$10$&$\rho_{22}^*$&$ 0.016$&[-0.084,0.096] &$41$\tabularnewline
$\rho_{13}^*$&$-0.147$&[-0.254,-0.063] &$ 8$&$\rho_{23}^*$&$-0.001$&[-0.141,0.106] &$64$\tabularnewline
$\rho_{14}^*$&$-0.199$&[-0.334,-0.097] &$ 6$&$\rho_{24}^*$&$ 0.037$&[-0.054,0.116] &$41$\tabularnewline
$\rho_{15}^*$&$-0.187$&[-0.343,-0.080] &$11$&$\rho_{25}^*$&$ 0.025$&[-0.096,0.114] &$51$\tabularnewline
$\rho_{16}^*$&$-0.166$&[-0.276,-0.079] &$ 7$&$\rho_{26}^*$&$-0.013$&[-0.109,0.070] &$31$\tabularnewline
$\rho_{17}^*$&$-0.225$&[-0.400,-0.111] &$ 8$&$\rho_{27}^*$&$-0.012$&[-0.125,0.081] &$45$\tabularnewline
$\rho_{18}^*$&$-0.173$&[-0.284,-0.084] &$ 8$&$\rho_{28}^*$&$ 0.021$&[-0.079,0.099] &$39$\tabularnewline
$\rho_{19}^*$&$-0.183$&[-0.337,-0.075] &$13$&$\rho_{29}^*$&$ 0.054$&[-0.066,0.143] &$55$\tabularnewline
\hline
\end{tabular}
}
\normalsize
\end{table}
\noindent
{\it Cholesky and spectral decompositions for computing $\mathbf{R}_t^{-1/2}$}. We also estimated our proposed models with $q=1$ and 2 using the Cholesky decomposition instead of the spectral decomposition. 
The estimation results using the Cholesky decomposition are very similar to those using the spectral decomposition (and hence are omitted) except for the parameters of the leverage effect. 
Table \ref{table:empirical_leverage_cholesky}  shows the estimation results for the correlation, $\rho_i^*$, with $q=1$. All posterior means are estimated to be negative and the posterior probability that $\rho_i^*$ is negative is greater than 0.975 for all $i$s. However, we note that the absolute values of $\rho_i^*$ are smaller than those in the model using the spectral decomposition. 
%
%
%
%
\begin{table}[H]
\footnotesize
\vspace{-2mm}
\centering
\caption{Posterior means, 95\% credible intervals and inefficiency factors for $\rho_i^*=Corr(z_{1t}, h_{i,t+1})$ where $\bm{z}_t=\mathbf{R}_t^{-1/2}\mathbf{V}_t^{-1/2}(\bm{y}_t-\bm{m}_t)$ and $q=1$. Cholesky decomposition is used.}
\vspace{2mm} 
\centering
\label{table:empirical_leverage_cholesky}
\scalebox{1.0}{
\begin{tabular}{lllr}
\hline\hline
\multicolumn{1}{l}{Par.}&\multicolumn{1}{c}{Mean}&\multicolumn{1}{c}{95\% interval}&\multicolumn{1}{c}{IF}\tabularnewline
\hline
$\rho_{1}^*$&$-0.170$&[-0.266,-0.092]&$7$\tabularnewline
$\rho_{2}^*$&$-0.118$&[-0.199,-0.049]&$4$\tabularnewline
$\rho_{3}^*$&$-0.091$&[-0.174,-0.021]&$8$\tabularnewline
$\rho_{4}^*$&$-0.118$&[-0.206,-0.045]&$7$\tabularnewline
$\rho_{5}^*$&$-0.123$&[-0.239,-0.044]&$7$\tabularnewline
$\rho_{6}^*$&$-0.119$&[-0.205,-0.048]&$7$\tabularnewline
$\rho_{7}^*$&$-0.141$&[-0.244,-0.061]&$9$\tabularnewline
$\rho_{8}^*$&$-0.144$&[-0.237,-0.067]&$10$\tabularnewline
$\rho_{9}^*$&$-0.148$&[-0.271,-0.058]&$8$\tabularnewline
\hline
\end{tabular}
}
\normalsize
\end{table}
\vspace{-4mm}
\begin{table}[H]
\footnotesize
\centering
\caption{Posterior means, 95\% credible intervals and inefficiency factors for $\rho_{1i}^*=Corr(z_{1t}, h_{i,t+1})$ and $\rho_{2i}^*=Corr(z_{2t}, h_{i,t+1})$ where $q=2$. Cholesky decomposition is used.}
\vspace{2mm} 
\label{table:empirical_leverage2_cholesky}
\scalebox{1.0}{
\begin{tabular}{lrcrlrcl}
\hline\hline
\multicolumn{1}{l}{Par.}&\multicolumn{1}{c}{Mean}&\multicolumn{1}{c}{95\% interval}&\multicolumn{1}{c}{IF} & 
\multicolumn{1}{l}{Par.}&\multicolumn{1}{c}{Mean}&\multicolumn{1}{c}{95\% interval}&\multicolumn{1}{c}{IF}\tabularnewline
\hline
$\rho_{11}^*$&-0.168&[-0.271,-0.087] &$7$ &$\rho_{21}^*$&-0.029 &[-0.089, 0.024] &$5$\tabularnewline
$\rho_{12}^*$&-0.132&[-0.230,-0.056] &$6$ &$\rho_{22}^*$&-0.113 &[-0.200,-0.043] &$7$\tabularnewline
$\rho_{13}^*$&-0.100&[-0.189,-0.029] &$7$ &$\rho_{23}^*$&-0.073 &[-0.147,-0.009] &$6$\tabularnewline
$\rho_{14}^*$&-0.120&[-0.212,-0.046] &$4$ &$\rho_{24}^*$&-0.019 &[-0.080, 0.036] &$7$\tabularnewline
$\rho_{15}^*$&-0.128&[-0.240,-0.045] &$9$ &$\rho_{25}^*$&-0.044 &[-0.117, 0.018] &$8$\tabularnewline
$\rho_{16}^*$&-0.125&[-0.215,-0.051] &$8$ &$\rho_{26}^*$&-0.028 &[-0.093, 0.029] &$5$\tabularnewline
$\rho_{17}^*$&-0.149&[-0.262,-0.064] &$7$ &$\rho_{27}^*$&-0.081 &[-0.160,-0.016] &$8$\tabularnewline
$\rho_{18}^*$&-0.145&[-0.239,-0.066] &$6$ &$\rho_{28}^*$&-0.011 &[-0.071, 0.042] &$5$\tabularnewline
$\rho_{19}^*$&-0.151&[-0.269,-0.058] &$10$&$\rho_{29}^*$&-0.003 &[-0.069, 0.052] &$7$\tabularnewline
\hline
\end{tabular}
}
\normalsize
\end{table}
\noindent
Table \ref{table:empirical_leverage2_cholesky} shows the estimation results for the correlations, $\rho_{1i}^*$ and $\rho_{2i}^*$, with $q=2$. The estimation results for $\rho_{1i}^*$  are similar to those for $\rho_i^*$ in Table \ref{table:empirical_leverage_cholesky}, but all posterior means of $\rho_{2i}^*$ are estimated to be negative, and the posterior probability that $\rho_{2i}^*$ is negative is greater than 0.975 for $i=2,3$ and $7$. This implies that we need to include more factors when we use the Cholesky decomposition. Since the number of factors $q$ depends on the order of the asset return, we have to find the order that minimizes $q$ for the parsimonious specification. Conversely, the spectral decomposition does not depend on the order of the asset returns and it is much faster at finding a parsimonious specification. We will compare these models using different decompositions with regard to their portfolio performances.
\subsection{Comparison of portfolio performances}
In order to compare the forecasting performance of our proposed models and other existing models,  we consider the minimum-variance portfolio strategy (see \cite{Han2006}).
We denote the conditional mean and the conditional covariance matrix of the stock return $\bm{y}_{t+1}$ given the information set $\calF_t$ at time $t$ as
\begin{eqnarray*}
\boldm_{t+1 \vert t} & \equiv & \E[\boldy_{t+1} \vert \calF_t]=\boldm_{t},
\quad
\boldSigma_{t+1 \vert t}  \equiv  \Var[\boldy_{t+1} \vert \calF_t]
= \boldV_{t+1}^{1/2}\boldR_{t+1}\boldV_{t+1}^{1/2} + \boldSigma_m.
\end{eqnarray*}
Let $r_{p,t+1}$ denote the portfolio return at time $t+1$. Further, we denote the conditional mean and conditional variance of $r_{p,t+1}$ given the information set $\calF_t$ at time $t$ by
\begin{eqnarray*}
\mu_{p,t+1}      &\equiv & 
\E[\boldw_t'\boldy_{t+1} + (1- \boldw_t'\boldone_p)r_f \vert \calF_t]
=\boldw_t'\boldm_{t+1 \vert t} + (1- \boldw_t'\boldone_p)r_f, \\
\sigma^2_{p,t+1} &\equiv & 
\Var[\boldw_t'\boldy_{t+1} + (1- \boldw_t'\boldone_p)r_f \vert \calF_t]
=\boldw_t' \boldSigma_{t+1 \vert t} \boldw_t,
\end{eqnarray*}
where $r_f$  is the risk free asset return, and $\boldw_t$ is a portfolio weight vector for the stock return $\bm{y}_{t+1}$. 
In the minimum-variance strategy, we minimize the conditional variance $\sigma^2_{p, t+1}$ for the target level $\mu_{p}^*$ of the conditional expected return $\mu_{p,t+1}$. Then the optimal weight $\boldw_t$ is given by
\begin{align*}
& \hat{\boldw}_t = \boldSigma_{t+1 \vert t}^{-1} (\boldm_{t+1 \vert t} - r_f \boldone_p) \frac{\mu_p^* - r_f}{\kappa_t}, 
\quad
\kappa_t = (\boldm_{t+1 \vert t} - r_f \boldone_p)' \boldSigma_{t+1 \vert t}^{-1}(\boldm_{t+1 \vert t} - r_f \boldone_p).
\end{align*}

\noindent
The portfolio performances are compared based on the rolling forecast:
\begin{enumerate}
\item[]\hspace{-7mm}Step 1. 
First, we estimate the parameters using the first $1742$ observations from February 1, 2001 to January 8, 2008 and forecast the mean, the volatility and the correlation of the multiple stock returns for January 9, 2008. We use them to obtain the optimal weights of the assets for the above portfolio strategies  and the federal funds (FF) rate is used for the risk free asset return $r_f$.

\item[]\hspace{-7mm}Step 2. 
Next, we drop the first observation (February 1, 2001) from the sample period and add the new observation (January 9, 2008). The new sample period is from February 2, 2001 to January 9, 2008. We estimate the parameters using these observations and forecast the mean, the volatility and the correlation for January 10, 2008. We use them to obtain the optimal weights in a similar manner.
\item[]\hspace{-7mm}Step 3.  We iterate these rolling forecasts until December 31, 2009 to obtain the 500 one-day ahead forecasts and the corresponding weights.
\end{enumerate}
To compute the optimal weight $\hat{\bm{w}}_t$, we also need the estimates of $\boldm_{t+1 \vert t}$ and $\boldSigma_{t+1 \vert t}$.
We let $N$ denote the number of MCMC iterations, and $(\theta^{(i)}, \{\boldh_t^{(i)} \}_{t=1}^{T}, \{\boldR_t^{(i)} \}_{t=1}^{T},  \{\boldm_t^{(i)} \}_{t=1}^{T} )$ denote the $i$-th MCMC sample ($i=1,\ldots,N$). 
Using $\boldm_{t+1 \vert t}^{(i)}, \boldV_{t+1\vert t}^{(i)}, \boldR_{t+1 \vert t}^{(i)}, \boldSigma_m^{(i)}$, we estimate $\boldm_{t+1 \vert t}$ and $\boldSigma_{t+1 \vert t}$ by 
\begin{align*}
&\hat{\boldm}_{t+1 \vert t} = \frac{1}{N} \sum_{i=1}^{N}
\boldm_{t+1 \vert t}^{(i)}, \quad
\hat{\boldSigma}_{t+1 \vert t} = \frac{1}{N} \sum_{i=1}^{N}\boldSigma_{t+1 \vert t}^{(i)} =
\frac{1}{N}\sum_{i=1}^{N} \left(\boldV_{t+1 \vert t}^{(i)1/2} \boldR_{t+1 \vert t}^{(i)}\boldV_{t+1 \vert t}^{(i)1/2} + \boldSigma_m^{(i)} \right).
\end{align*}
In our empirical study, we set $N=1500$ and we discard  $500$ samples as the burn-in period for each MCMC rolling estimation (Steps 2 and 3)\footnote{The number of samples being discarded as the burn-in period is sufficient after we obtain the MCMC posterior samples from the previous sample period since we use the posterior means of the parameters and latent variables for the initial values of the next MCMC runs.}. We compare the following multivariate stochastic volatility models as follows. 
\small
\begin{enumerate}
\item MSV model:  Basic multivariate stochastic volatility model without leverage, realized variances and correlations.

\item CRSV model: Cholesky realized stochastic volatility model with leverage proposed in \cite{ShirotaOmoriLopesPiao(17)}

\item MRSV model: Multivariate stochastic volatility model without leverage and with realized variances and pairwise realized correlations. 

\item MRSV-L1-C model: Multivariate stochastic volatility model with leverage, realized variances and pairwise realized correlations.  The parsimonious specification is assumed to model the leverage effect,  $\mathbf{\Lambda}=[\bm{\lambda}_1,\bm{0},\ldots,\bm{0}]$ with $q=1$. The Cholesky decomposition is used to compute $\mathbf{R}_t^{-1/2}$.

\item MRSV-L2-C model: Multivariate stochastic volatility model with leverage, realized variances and pairwise realized correlations.  The parsimonious specification is assumed to model the leverage effect,  $\mathbf{\Lambda}=[\bm{\lambda}_1,\bm{\lambda}_2, \bm{0},\ldots,\bm{0}]$ with $q=2$. The Cholesky decomposition is used to compute $\mathbf{R}_t^{-1/2}$.

\item MRSV-L1-S model: Multivariate stochastic volatility model with leverage, realized variances and pairwise realized correlations.  The parsimonious specification is assumed to model the leverage effect,  $\mathbf{\Lambda}=[\bm{\lambda}_1,\bm{0},\ldots,\bm{0}]$ with $q=1$. The spectral decomposition is used to compute $\mathbf{R}_t^{-1/2}$. 

\item MRSV-L1-S (constant mean) model: The mean vector $\bm{m}_t$ of the return is assumed to be constant in the MRSV-L1-S model.

\item DCC-GARCH model: DCC-GARCH model proposed in \cite{Engle(02)}\footnote{The parameters are estimated by the maximum likelihood method.}.

\item HEAVY model: a scalar HEAVY model proposed in \cite{noureldin2012multivariate} for each element of the spectral decomposition of the realized covariance matrix.  \footnote{The parameters are estimated by the two step estimation.
The mean of the return is estimated by the corresponding sample mean during the sample period.}.

\item HAR model: HAR model proposed in \cite{Corsi(09)}\footnote{The mean of the return is estimated by the corresponding sample mean during the sample period.}.
\item Equally weighted portfolio model: The weights of the assets are fixed to be equal in the model.
\end{enumerate}
\normalsize
%
%
%
%
%
%
\vspace{-4mm}
\begin{table}[H]
\footnotesize
\centering
\caption{The cumulative values of realized objective functions.}
\vspace{2mm}
\label{table:portfolio_cum_obj}
\scalebox{1.0}{
\begin{tabular}{lrrr}
\hline\hline
~~&$\mu_p^* = 0.004$&$\mu_p^* = 0.01$&$\mu_p^* = 0.1$\tabularnewline
\hline
~~MSV               &$1.172$&$6.536$&$1184$\tabularnewline
~~CRSV              &$0.748$&$4.448$&$ 730$\tabularnewline
~~MRSV              &$0.526$&$2.943$&$ 510$\tabularnewline
~~MRSV-L1-C         &$0.272$&$1.601$&$ 262$\tabularnewline
~~MRSV-L2-C         &$0.264$&$1.552$&$ 255$\tabularnewline
~~MRSV-L1-S         &${\bf 0.249}$&${\bf 1.430}$&${\bf 232}$\tabularnewline
~~MRSV-L1-S (constant mean) &$0.568$&$3.032$&$ 543$\tabularnewline
~~DCC-GARCH         &$2.662$&$11.962$&$ 2537$\tabularnewline
~~HEAVY             &$2.888$&$14.824$&$ 3058$\tabularnewline
~~HAR               &$2.537$&$13.110$&$ 2680$\tabularnewline
~~Equal weight      &$1425$ &$1425$ &$ 1425$\tabularnewline
\hline\hline
\end{tabular}
}
\footnotesize
\begin{flushleft}
*Bold figures indicates the optimal values. The cumulative variances are computed as $\sum_{t=1742}^{2241}\hat{\bm{\omega}}_t'\mathbf{\Sigma}_{t+1}\hat{\bm{\omega}}_t$ where $\mathbf{\Sigma}_{t+1}$ is evaluated using the realized covariance at time $t+1$. The results for CRSV model are reproduced from Shirota {\it et al.} (2017).
\end{flushleft}
\normalsize
\end{table}
\vspace{-6mm}
\noindent
{\it Cumulative realized objective functions.}
Table \ref{table:portfolio_cum_obj} shows the cumulative values of the realized objective functions.
The MRSV-L1-S model outperforms the other models. Among the MRSV models, the models with leverage outperform the models without leverage, thereby indicating the existence and the importance of the leverage effect. If we assume a constant mean for $\bm{y}_t$, the performance becomes poor in this prediction period, which implies that the random walk process is more flexible for describing the dynamics of the mean of the return vector.  The MRSV-L2-C model outperforms the MRSV-L1-C model, but its performance is not as good as that of the MRSV-L1-S model. We could improve the performance of the MRSV models using the Cholesky decomposition by changing the order of the assets or increasing the number of nonzero columns of $\mathbf{\Lambda}$, but it is more efficient to use the spectral decomposition to compute $\mathbf{R}_t^{-1/2}$. 
Finally, in comparison with the DCC-GARCH model, the HEAVY model, the HAR model and  the equally weighted model, we found that the classes of the MSV and MRSV models perform much better.
%
%
\begin{figure}[H]
\vspace{-3mm}
\centering
\caption{Time series plot of the portfolio weights n MRSV-L1-S: $\mu_p^* = 0.1$.\label{figure:portfolio_weight_minvar}}
\includegraphics[width = 0.7\linewidth, angle = 0]{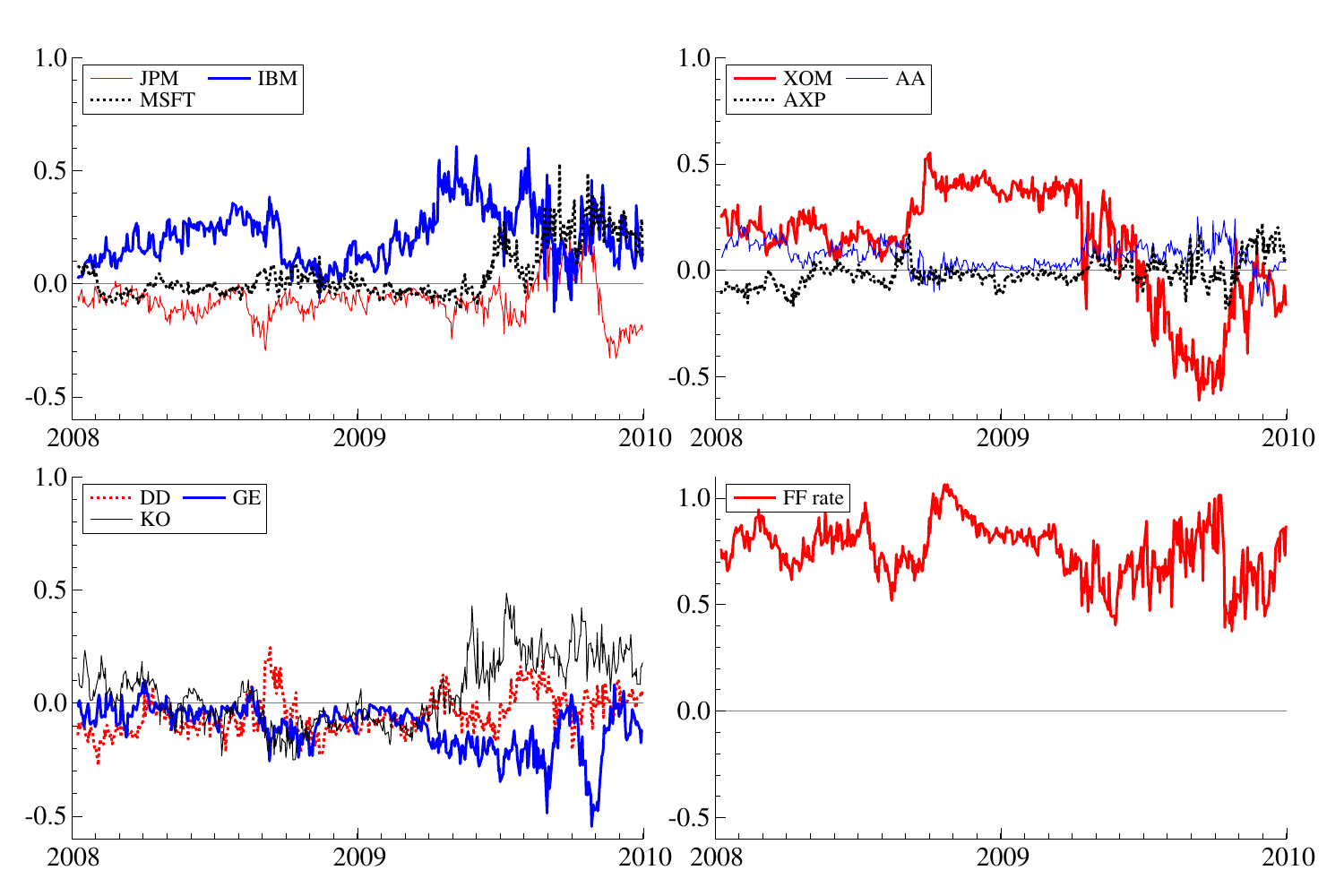}
\end{figure}
%
\noindent
{\it Time series plots of the portfolio weights}.
Figure \ref{figure:portfolio_weight_minvar} shows the time series plots of the portfolio weights in the MRSV-L1-S. The weights for Exxon Mobil are large among all stocks, but towards the end of the period, the weights for IBM and Microsoft tend to become large. However, the weights for the FF rate ($1-\sum_{i=1}^9w_{it}$) are the largest throughout the forecasting period.\\

\noindent
{\it Comparison of performances before and after the financial crisis}.
In order to illustrate the portfolio performances of the models (excluding the CRSV model) in more detail before and after the financial crisis, we divide the forecasting periods into two subperiods: (1) Jan 9, 2008 - July 31, 2008 and (2) Aug 1, 2008 - Dec 31, 2009. As shown in Table \ref{table:portfolio_cum_obj_sub}, in both subperiods (1) and (2), the portfolio performances  are similar to those that we found in the whole period. 
\begin{table}[H]
\footnotesize
\centering
\caption{The cumulative values of realized objective functions in two subperiods.}
\vspace{2mm}
\label{table:portfolio_cum_obj_sub}
\scalebox{1.0}{
\begin{tabular}{lrrr}
\hline\hline
\multicolumn{4}{l}{\bfseries (1) Jan 9, 2008 - July 31, 2008}\tabularnewline
~~&$\mu_p^* = 0.004$&$\mu_p^* = 0.01$&$\mu_p^* = 0.1$\tabularnewline
\hline
~~MSV               &$0.436$&$0.357$&$ 253$\tabularnewline
~~MRSV              &$0.175$&$0.110$&$  87$\tabularnewline
~~MRSV-L1-C         &$0.072$&$0.042$&$  33$\tabularnewline
~~MRSV-L2-C         &$0.073$&$0.044$&$  34$\tabularnewline
~~MRSV-L1-S         &${\bf 0.072}$&${\bf 0.041}$&${\bf  33}$\tabularnewline
~~MRSV-L1-S (constant mean) &$0.213$&$0.139$&$ 107$\tabularnewline
~~DCC-GARCH         &$1.265$&$1.074$&$ 769$\tabularnewline
~~HEAVY             &$1.099$&$1.250$&$ 810$\tabularnewline
~~HAR               &$0.958$&$1.038$&$ 683$\tabularnewline
~~Equal weight      &$  176$&$  176$&$ 176$\tabularnewline
\hline
\multicolumn{4}{l}{\bfseries  (2) Aug 1, 2008 - Dec 31, 2009}\tabularnewline
~~&$\mu_p^* = 0.004$&$\mu_p^* = 0.01$&$\mu_p^* = 0.1$\tabularnewline
\hline
~~MSV               &$0.736$&$6.179$&$ 932$\tabularnewline
~~MRSV              &$0.351$&$2.833$&$ 423$\tabularnewline
~~MRSV-L1-C         &$0.200$&$1.559$&$ 228$\tabularnewline
~~MRSV-L2-C         &$0.191$&$1.508$&$ 220$\tabularnewline
~~MRSV-L1-S         &${\bf 0.177}$&${\bf 1.389}$&${\bf 199}$\tabularnewline
~~MRSV-L1-S (constant mean) &$0.354$&$2.893$&$ 436$\tabularnewline
~~DCC-GARCH         &$1.397$&$10.889$&$ 1768$\tabularnewline
~~HEAVY             &$1.789$&$13.574$&$ 2248$\tabularnewline
~~HAR               &$1.579$&$12.072$&$ 1997$\tabularnewline
~~Equal weight      &$ 1248$&$ 1248$&$ 1248$\tabularnewline
\hline\hline
\normalsize
\end{tabular}
}
\end{table}
\vspace{-6mm}
\noindent
{\it Remark} 4.
As suggested by an anonymous referee and the Editor, we conducted a predictive ability test based on \cite{GiacominiWhite(06)} in order to investigate whether the realized objective function of each model is significantly different from that of the MRSV-L1-S in Tables 9 and 10. We found that the differences are all significant except for the MRSV-L1-C models ($\mu_p^*=0.004,0.01,0.1$) and 
MRSV-L2-C models ($\mu_p^*=0.004,0.1$) in the subperiod (1) (which is before the financial crisis).
\section{Conclusions}
The multivariate SV model with flexible dynamic correlation structures that uses the Markov chain Monte Carlo estimation method is proposed.  By making full use of the realized variances and realized pairwise correlations, we obtain stable parameter estimates where the covariance matrices are guaranteed to be positive definite. The spectral decomposition is used for the correlation matrices in order to avoid the arbitrariness of the ordering of  asset returns. The parsimonious specification for the leverage effect is also proposed.
Our models are applied to the daily returns of nine U.S. stocks with their realized volatilities and pairwise realized correlations and are shown to outperform the existing models with regard to  portfolio optimizations under a minimum-variance strategy.\vspace{2mm}\\
\noindent
{\bf \large Acknowledgements}\\
We thank anonymous referees, the Editor, John Maheu, Hideo Kozumi, Masahiko Sagae and Shuji Tanaka for providing useful comments and discussions. The computational results were obtained by using Ox version 7 (\cite{Doornik(06)}). This work was supported by JSPS KAKENHI Grant Numbers 25245035, 26245028.%
\small
\section*{Appendix}
\appendix
\section{MCMC algorithm}
\subsection{MRSV model without the leverage effect}
\label{sec:mcmc_wo_leverage}
\subsubsection{Joint posterior density}
The joint posterior probability density function is given by
\begin{align}
\nonumber
	&\pi(\bvec{\theta},\bvec{g},\bvec{h}, \boldm \vert \bvec{w},\bvec{x},\bvec{y}) \\
\nonumber
	&\propto \prod^{T}_{t=1} \dtm{ \boldV_t^{1/2}\boldR_t\boldV_t^{1/2} }^{-1/2} \exp\left\{ -\frac{1}{2}(\boldy_t - \boldm_t)'(\boldV_t^{1/2}\boldR_t\boldV_t^{1/2})^{-1} (\boldy_t - \boldm_t)\right\} \\
\nonumber
	& \quad \times |\boldOmega_0|^{-1/2}\exp\left\{-\frac{1}{2}(\boldh_1-\bvec{\mu})'\boldOmega_0^{-1} (\boldh_1-\bvec{\mu}) \right\}  \\
\nonumber
	& \quad \times \prod^{T-1}_{t=1}|\bvec{\Omega}|^{-1/2}\exp\left[-\frac{1}{2}\{\boldh_{t+1}-(\mathbf{I}_p-\bvec{\Phi})\bvec{\mu} - \bvec{\Phi} \boldh_t \}' \bvec{\Omega}^{-1} \{\boldh_{t+1}-(\mathbf{I}_p-\bvec{\Phi})\bvec{\mu} - \bvec{\Phi} \boldh_t \} \right] \\
\nonumber
	& \quad \times \prod_{t=1}^{T}
	|\boldSigma_u|^{-1/2}\exp\left\{-\frac{1}{2}(\boldx_t-\bvec{\xi}-\boldh_t)'\boldSigma_u^{-1} (\boldx_t-\bvec{\xi}-\boldh_t) \right\} \\
\nonumber
	& \quad \times \prod_{i>j}^{p} \sigma_{\zeta,ij}^{-1} \exp\left(-\frac{g_{ij,1}^2}{2\kappa\sigma^2_{\zeta,ij}}\right) \prod_{t=1}^{T-1} \sigma_{\zeta,ij}^{-1} \exp\left\{-\frac{(g_{ij,t+1}-g_{ij,t})^2}{2\sigma^2_{\zeta,ij}}\right\} \\
\nonumber
	& \quad \times \prod_{i>j}^{p} \prod\limits_{t=1}^{T}\sigma_{v,ij}^{-1} \exp\left\{-\frac{(w_{ij,t}-\delta_{ij}-g_{ij,t})^2}{2\sigma^2_{v,ij}}\right\} \\
	& \quad \times \prod_{i=1}^{p} \sigma_{m,i}^{-1} \exp\left(-\frac{m_{i,1}^2}{2\kappa \sigma^2_{m,i}}\right) 
\prod\limits_{t=1}^{T-1}\sigma_{m,i}^{-1} \exp\left\{-\frac{(m_{i,t+1}-m_{i,t})^2}{2\sigma^2_{m,i}}\right\} 
\times \pi(\boldtheta),
\end{align}
where $\boldSigma_u = \diag(\sigma_{u,1}^2,\ldots, \sigma_{u,p}^2)$ and $\pi(\boldtheta)$ is a prior probability density function of parameters.
\subsubsection{Generation of $\bvec{\phi}$}
\label{sec:mcmc_phi}
It can be shown that the conditional posterior probability density function of $\boldphi$ is
	\begin{align}
		\pi(\bvec{\phi}|\cdot)
		&\propto
		k(\bvec{\phi})
		\times
		\exp\left(
		-\frac{1}{2}
		(\bvec{\phi}-\bvec{\mu_{\bvec{\phi}}})'\bvec{\Sigma}_{\bvec{\phi}}^{-1}
		(\bvec{\phi}-\bvec{\mu_{\bvec{\phi}}})
		\right)
		\times
		I\left\{ |\phi_i| <1,  i = 1,\ldots,p \right\}
		,
	\end{align}
where $I(B)$ is an indicator function such that $I(B)=1$ if $B$ is true and 0 otherwise, 
	\begin{align}
		&k(\bvec{\phi})
		 = |\bvec{\Omega}_0|^{-1/2}
		\prod_{i=1}^{p} \left(\frac{1+\phi_i}{2}\right)^{a-1}\left(\frac{1-\phi_i}{2}\right)^{b-1}
		\exp\left(
		-\frac{1}{2}
		(\bvec{h}_1 -\bvec{\mu})'\bvec{\Omega}_0^{-1}(\bvec{h}_1 -\bvec{\mu})
		\right),
		\\
		&\bvec{\mu}_{\bvec{\phi}} = \bvec{\Sigma}_{\bvec{\phi}}\bvec{b},
		\quad \bvec{\Sigma}_{\bvec{\phi}}^{-1} = \bvec{\Omega}^{-1}\odot\mathbf{A}, \\
		&\mathbf{A} = \sum_{t=1}^{T-1}(\bvec{h}_t -\bvec{\mu})(\boldh_t-\bvec{\mu})',
		\quad \bvec{b} = \text{diagonal}\left\{
				\sum_{t=1}^{T-1}(\bvec{h}_t - \bvec{\mu})(\bvec{h}_{t+1} - \bvec{\mu})' \bvec{\Omega}^{-1}
				\right\}
		,
	\end{align}
	$\odot$ is Hadamard product, and $\text{diagonal}(\mathbf{B})$ denotes a column vector with diagonal elements of $\mathbf{B}$.
	We propose a candidate $\bvec{\phi}^{\dagger} \sim \text{TN}_R(\bvec{\mu}_{\bvec{\phi}},\bvec{\Sigma}_{\bvec{\phi}})$, where $R = \{\bvec{\phi}:|\phi_i| <1 , i = 1,\ldots,p\}$, and accept it with probability  $\min\{1,k(\bvec{\phi}^{\dagger})/k(\bvec{\phi})\}$.

\subsubsection{Generation of $\bvec{\mu},\bvec{\xi},\bvec{\delta}$}
\label{sec:mcmc_mu_xi_delta}
The $\bvec{\mu}, \bvec{\xi}$ and $\bvec{\delta}$ are conditionally independent and we generate them from the following normal distributions:
\begin{align}
\bvec{\mu}|\cdot \sim N(\bvec{\tilde{m}_\mu},\bvec{\tilde{\Omega}_\mu}),
\quad\bvec{\xi}|\cdot \sim N(\bvec{\tilde{m}_\xi},\bvec{\tilde{\Sigma}_\xi}),
\quad\bvec{\delta}|\cdot\sim N(\bvec{\tilde{m}_\delta},\bvec{\tilde{\Sigma}_\delta}),
\end{align}
where
\begin{align}
	&\bvec{\tilde{m}_\mu}=\bvec{\tilde{\Omega}_\mu}\left[s_\mu^{-2}\bvec{m_\mu}+\boldOmega_0^{-1}\boldh_1+(\boldI_p-\bvec{\Phi})\bvec{\Omega}^{-1}\sum_{t=1}^{T-1}(\boldh_{t+1}-\bvec{\Phi}\boldh_t) \right], \\
	&\bvec{\tilde{\Omega}_\mu}= \left[s_\mu^{-2} \boldI_p+\boldOmega_0^{-1} + (T-1)(\boldI_p-\bvec{\Phi}) \bvec{\Omega}^{-1} (\boldI_p-\bvec{\Phi}) \right]^{-1}, \\
	&\tilde{\boldm}_\xi = \tilde{\boldSigma}_\xi  \left[ s_\xi^{-2}\boldm_\xi+\boldSigma^{-1}_u\sum_{t=1}^{T}(\boldx_t-\boldh_t)\right],
	\quad \tilde{\boldSigma}_\xi = (s_\xi^{-2} \boldI_p+T \boldSigma_u^{-1})^{-1}, \\
	&\tilde{\boldm}_\delta = \tilde{\boldSigma}_\delta
	\left[
	s_\delta^{-2}\boldm_\delta+\boldSigma^{-1}_v\sum_{t=1}^{T}(\boldw_t-\boldg_t) \right],
	\quad \tilde{\boldSigma}_\delta = (s_\delta^{-2} \boldI_p+T \boldSigma_v^{-1})^{-1}
	.
\end{align}
\subsubsection{Generation of  $(\boldsigma_{u}^2,\boldsigma_{v}^2,\boldsigma_{\zeta}^2,\mathbf{\Sigma}_m)$}
The $\sigma_{u,i}^2,\sigma_{v,ij}^2,\sigma_{\zeta,ij}^2,\sigma_{m,i}^2$ are conditionally independent and we generate them from inverse gamma distributions:
\begin{align*}
& \sigma_{u,i}^2\sim \text{IG}\left(\frac{\tilde{n}_{ui}}{2},\frac{\tilde{d}_{ui}}{2}\right),\quad\sigma_{v,ij}^2\sim \text{IG}\left(\frac{\tilde{n}_{v,ij}}{2},\frac{\tilde{d}_{v,ij}}{2}\right),
\quad
\sigma_{\zeta,ij}^2\sim \text{IG}\left(\frac{\tilde{n}_{\zeta,ij}}{2},\frac{\tilde{d}_{\zeta,ij}}{2}\right),
\quad \sigma_{m,i}^2\sim \text{IG}\left(\frac{\tilde{n}_{mi}}{2},\frac{\tilde{d}_{mi}}{2}\right),
\end{align*}
where
\begin{align}
	&\tilde{n}_{ui}=n_u+T, \quad \tilde{d}_{ui}=d_u+\sum_{t=1}^{T}(x_{it}-\xi_i-h_{it})^2,\\
	&\tilde{n}_{v,ij}=n_v+T, \quad \tilde{d}_{v,ij}=d_v+\sum_{t=1}^{T}(w_{ij,t}-\delta_{ij}-g_{ij,t})^2,\\
	&\tilde{n}_{\zeta,ij}=n_\zeta+T, \quad \tilde{d}_{\zeta,ij}=d_\zeta + \kappa^{-1}g_{ij,1}^2+\sum_{t=1}^{T-1}(g_{ij,t+1}-g_{ij,t})^2,\\
	&\tilde{n}_{mi}=n_m+T, \quad \tilde{d}_{mi}=d_m + \kappa^{-1}m_{i1}^2+\sum_{t=1}^{T-1}(m_{i,t+1}-m_{it})^2,
\end{align}
for $i,j=1,\ldots,p$ $(i>j)$.
\subsubsection{Generation of $\bvec{\Omega}$}
The conditional posterior probability density function of $\bvec{\Omega}$ is 
\begin{align}
	\pi(\bvec{\Omega}|\bvec{h},\bvec{\mu},\bvec{\phi}) \propto m(\bvec{\Omega}) \times 
|\mathbf{\Omega}|^{-(\tilde{\nu}+p+1)/2}
\exp \left\{-\frac{1}{2}\tr \left(\mathbf{\Omega}^{-1}\tilde{\mathbf{S}}\right)\right\},
\end{align}
where
\begin{eqnarray}
m(\bvec{\Omega})
	& = & |\boldOmega_0|^{-1/2}\exp\left(-\frac{1}{2}(\boldh_1-\bvec{\mu})'\boldOmega_0^{-1} (\boldh_1-\bvec{\mu}) \right), \quad \tilde{\nu}=\nu+T-1,
\\
\tilde{\bvec{S}} & = &\bvec{S}+\sum_{t=1}^{T-1}\{\boldh_{t+1}-(\boldI_p-\bvec{\Phi})\bvec{\mu} - \bvec{\Phi} \boldh_t \}\{\boldh_{t+1}-(\boldI_p-\bvec{\Phi})\bvec{\mu} - \bvec{\Phi} \boldh_t \}'.
\end{eqnarray}
We propose a candidate $\bvec{\Omega}^{\dagger} \sim \IW(\tilde{\nu},\tilde{\bvec{S}})$, and accept it with probability $\min \{1, m(\bvec{\Omega}^{\dagger})/m(\bvec{\Omega})\}$.
\subsubsection{Generation of $\mathbf{\Psi}$}
As a prior distribution of $\mathbf{\Psi}$, we assume $\mathbf{\Psi}\sim \text{IW}(\nu_{\psi},\mathbf{S}_{\psi})$.  Then the conditional posterior probability density function of $\boldPsi$ is
\begin{align}
	\pi(\bvec{\Psi}|\cdot) \propto m(\bvec{\Psi}) \times
	|\mathbf{\Psi}|^{-(\tilde{\nu}_{\psi}+p+1)/2} \exp \left\{-\frac{1}{2}\tr \left(\mathbf{\Psi}^{-1}\tilde{\mathbf{S}}_\psi\right)\right\},
\end{align}
where
\begin{eqnarray}
m(\bvec{\Psi})
	&=&|\boldOmega_0|^{-1/2}\exp\left(-\frac{1}{2}(\boldh_1-\bvec{\mu})'\boldOmega_0^{-1} (\boldh_1-\bvec{\mu}) \right), \\
\tilde{\nu}_\psi 
	&=& \nu_\psi+p+T-1,\\
\nonumber
\tilde{\mathbf{S}}_\psi
	&= &  \mathbf{S}_\psi +(\mathbf{\Lambda} - \boldM_0)\mathbf{\Gamma}_0^{-1} (\mathbf{\Lambda} - \boldM_0)' \\
\nonumber
	& &\hspace{5mm}+\sum_{t=1}^{T-1}
	\left\{\boldh_{t+1} - \boldmu - \boldPhi(\boldh_t - \boldmu) - \boldLambda\boldR_t^{-1/2} \boldV_t^{-1/2} (\boldy_t - \boldm_t)\right\} \\
	& &\hspace{3cm}\left\{\boldh_{t+1} - \boldmu - \boldPhi(\boldh_t - \boldmu) - \boldLambda\boldR_t^{-1/2} \boldV_t^{-1/2} (\boldy_t - \boldm_t)\right\}'
	.
\end{eqnarray}
Thus we propose a candidate $\mathbf{\Psi}^{\dagger}$ from $\IW(\tilde{\nu}_{\psi},\tilde{\mathbf{S}}_{\psi})$, and accept it with probability \\
$\min\{1, m(\mathbf{\Psi}^{\dagger})/m(\mathbf{\Psi})\}$.
\subsubsection{Generation of $\boldmu$}
Noting that
\begin{align}
&\E[\boldy_t  \vert \boldh_{t}] = \boldm_t, \quad \Var[\boldy_t \vert \boldh_{t}] = \boldV_t^{1/2}\boldR_t\boldV_t^{1/2} \equiv\boldGamma_t , 
\end{align}
it can be shown that the conditional posterior distribution of $\bm{m}_t$ is the same as that of 
the following linear Gaussian state space model:
\begin{eqnarray}
\label{eq:ssm_mt_1}
\boldy_t &=& \boldm_t + \hat{\boldeps}_t , \quad \hat{\boldeps}_t \sim \Normal(\boldzero, \boldGamma_t), \\
\label{eq:ssm_mt_2}
\boldm_{t+1} &=& \boldm_t + \boldnu_t, \quad \boldnu_t \sim \Normal(\boldzero, \boldSigma_m),
\end{eqnarray}
where $\hat{\boldeps}_t$ and $\boldnu_t$ are independent. Thus we generate $\bm{m}$ simultaneously at one time using a simulation smoother (e.g. \cite{deJongShephard(95)}, \cite{DurbinKoopman(02)}).
\subsection{MRSV model with the leverage effect}
\label{sec:mcmc_w_leverage}
We need to modify the sampling procedures of $\bm{g}, \bm{h}, \bm{m}, \bm{\phi}$ and $\bm{\mu}$ for the model with the leverage effect.
Generations of other parameters are the same as in the previous section. 

\subsubsection{Generation of $\bm{g}_t$}
We only need to modify (\ref{eq:r_g}) in Section \ref{sec:mcmc_gt} as follows.
\begin{align}
\nonumber
	r(g_{ij,t})
	=
	&-\frac{1}{2}\log\dtm{\boldR_t}  -\frac{1}{2}(\boldy_t - \boldm_t)'(\boldV_t^{1/2}\boldR_t\boldV_t^{1/2})^{-1}(\boldy_t - \boldm_t)\\
	&-\frac{1}{2} \boldy'_t \boldV^{-1/2} \boldR_t^{-1/2'} \boldLambda' \boldPsi^{-1}\boldLambda  \boldR_t^{-1/2} \boldV_t^{-1/2}(\boldy_t - \boldm_t)
	+ \boldy'_t \boldV_t^{-1/2} \boldR_t^{-1/2'}\boldLambda' \boldPsi^{-1}\boldeta_t,
\end{align}
for $t=1,\ldots,T-1$ where $\bm{\eta}_t=\bm{h}_{t+1}-\bm{\mu}-\mathbf{\Phi}(\bm{h}_t-\bm{\mu})$.

\subsubsection{Generation of $\boldh_t$}
The conditional posterior probability density function of $\boldh_t$ is given by
\begin{align}
\pi (\boldh_t \vert \cdot)
	\propto
	\exp\left[
	-\frac{1}{2}(\boldh_t - \bm{m}_{t*})'\boldOmega_{t*}^{-1}(\boldh_t - \bm{m}_{t*}) +l(\boldh_t)
	\right],
\end{align}
where
\begin{eqnarray}
\nonumber
\lefteqn{l(\boldh_t)} && \\
&=& 
\left\{
\begin{array}{ll}
	-\frac{1}{2} (\boldy_t - \boldm_t)' \boldV^{-1/2} \boldR_t^{-1} \boldV_t^{-1/2} (\boldy_t - \boldm_t)
	-\frac{1}{2} (\boldy_t-\bm{m}_t)' \boldV_t^{-1/2} \boldR_t^{-1/2'} \boldLambda' \boldPsi^{-1}
& \\
	\hspace{3mm}\times \left\{
	\boldLambda  \boldR_t^{-1/2} \boldV_t^{-1/2}(\boldy_t - \boldm_t) 
	-2(\boldh_{t+1} - (\boldI - \boldPhi) \boldmu - \boldPhi \boldh_t)
	\right\}, & \\
	&  \hspace{-3cm}t=1,\ldots,T-1, 
\\
	-\frac{1}{2} (\boldy_T - \boldm_T)' \boldV^{-1/2}_T \boldR_T^{-1} \boldV_T^{-1/2} (\boldy_T - \boldm_T), & \hspace{-1.2cm}t=T,
\end{array}
\right.
\end{eqnarray}
and
\begin{eqnarray}
\bm{m}_{t*} & = & \left\{
	\begin{array}{ll}
	\mathbf{\Psi}_{1*}
	\left[
		\boldOmega_0^{-1} \boldmu
		+ \boldPhi \boldPsi^{-1}
		\left\{
			\boldh_{2} - (\boldI_p - \boldPhi)\boldmu
		\right\}
			+ \boldSigma_u^{-1}
			(\boldx_1 - \boldxi)
			- \frac{1}{2} \boldone_p
	\right], & t=1,\\
	\boldPsi_{t*}
	\bigg[
		\boldPsi^{-1} 
		\left\{
		\boldLambda \boldR_{t-1}^{-1/2} \boldV_{t-1}^{-1/2} (\boldy_{t-1} - \boldm_{t-1})+(\boldI_p - \boldPhi)\boldmu + \boldPhi \boldh_{t-1}
		\right\}
	 &   \\
		\hspace{2cm}
		+ \boldPhi \boldPsi^{-1}
		\left\{
		\boldh_{t+1} - (\boldI_p - \boldPhi)\boldmu
		\right\}
		+ \boldSigma_u^{-1}(\boldx_t - \boldxi) - \frac{1}{2} \boldone_p
	\bigg], & \\
	& \hspace{-1.7cm} t=2,\ldots,T-1,\\
	\boldPsi_{T*}
	\left[
		\boldPsi^{-1} 
		\left\{
		\boldLambda \boldR_{T-1}^{-1/2} \boldV_{T-1}^{-1/2} (\boldy_{T-1} - \boldm_{T-1})+(\boldI_p - \boldPhi)\boldmu + \boldPhi \boldh_{T-1}
		\right\}
	\right. &   \\
	\left.	\hspace{2cm}
		+ \boldSigma_u^{-1}(\boldx_T - \boldxi) - \frac{1}{2} \boldone_p
	\right], & t=T,\\
	\end{array}
\right.
\\
\mathbf{\Psi}_{t*} & = & \left\{
	\begin{array}{ll}
	\left[
		\boldOmega_0^{-1}
		+ \boldPhi \boldPsi^{-1} \boldPhi
		+ \boldSigma_u ^{-1}
	\right]^{-1}, & t=1,\\
	\left[
		\boldPsi^{-1} + \boldPhi \boldPsi^{-1} \boldPhi + \boldSigma_u ^{-1}
	\right]^{-1}, & t=2,\ldots,T-1,\\
	\left[
		\boldPsi^{-1} + \boldSigma_u ^{-1}
	\right]^{-1}, & t=T.
	\end{array}
\right.
\end{eqnarray}
We generate a candidate $\bm{h}_t^{\dagger}$ from $\Normal(\bm{m}_{t*}, \boldOmega_{t*})$, and accept it with probability $\min\{1,\exp(l(\bm{h}_t^{\dagger}) - l(\bm{h}_t))\}$.

\subsubsection{Generation of $\boldm$}
Noting that
\begin{eqnarray}
E[\boldy_t  \vert \boldh_{t}, \boldh_{t+1},\bm{\theta}] 
	&=& \boldm_t + \mathbf{V}_t^{1/2}\mathbf{R}_t^{-1/2}\mathbf{\Lambda}'
	(\mathbf{\Psi}+\mathbf{\Lambda}\mathbf{\Lambda}')^{-1}\{\bm{h}_{t+1}-\bm{\mu}-\mathbf{\Phi}(\bm{h}_t-\bm{\mu})\},
\\
\Var[\boldy_t \vert \boldh_{t}, \boldh_{t+1},\bm{\theta}] 
	&=& \boldV_t ^{1/2}\boldR_t\boldV_t^{-1/2}-
\mathbf{V}_t^{1/2}\mathbf{R}_t^{-1/2}\mathbf{\Lambda}'(\mathbf{\Psi}+\mathbf{\Lambda}\mathbf{\Lambda}')^{-1}\mathbf{\Lambda}\mathbf{R}_t^{-1/2\prime}\mathbf{V}_t^{1/2}\equiv \mathbf{\Gamma}_t,
\label{eq:Gamma_t_leverage}
\hspace{6mm}\mbox{}
\end{eqnarray}
we define
\begin{align}
&\hat{\boldy}_t = \boldy_t - \mathbf{V}_t^{1/2}\mathbf{R}_t^{-1/2\prime}\mathbf{\Lambda}'
	(\mathbf{\Psi}+\mathbf{\Lambda}\mathbf{\Lambda}')^{-1}\{\bm{h}_{t+1}-\bm{\mu}-\mathbf{\Phi}(\bm{h}_t-\bm{\mu})\},
\end{align}
and consider the linear Gaussian state space model (\ref{eq:ssm_mt_1})
 and (\ref{eq:ssm_mt_2}) with $\mathbf{\Gamma}_t$ in (\ref{eq:Gamma_t_leverage}). We generate $\boldm$ simultaneously using a simulation smoother.

\subsubsection{Generation of $\bm{\phi}$}
In Section \ref{sec:mcmc_phi}, we replace $\bvec{\mu}_{\bvec{\phi}}$ and $\bvec{\Sigma}_{\bvec{\phi}}$ as follows.
\begin{align*}
\bvec{\mu}_{\bvec{\phi}} = \bvec{\Sigma}_{\bvec{\phi}}\bvec{b}, \quad
\bvec{\Sigma}_{\bvec{\phi}}^{-1} = \bvec{\Psi}^{-1}\odot\mathbf{A},
\end{align*}
where
\begin{align*}
\boldA = \sum_{t=1}^{T-1}(\bvec{h}_t -\bvec{\mu})(\boldh_t-\bvec{\mu})',
\quad
\bm{b}= \text{diagonal}\left[
			\sum_{t=1}^{T-1}
			(\bvec{h}_t - \bvec{\mu})
			\left\{\bvec{h}_{t+1} - \bvec{\mu} 
				-\boldLambda\boldR_t^{-1/2}\boldV_t^{-1/2}(\boldy_t - \boldm_t)
			\right\}' \bvec{\Psi}^{-1}
			\right]
	.
\end{align*}

\subsubsection{Generation of $\mathbf{\Psi}$}
As the prior distribution of $\mathbf{\Lambda}$ is changed, the conditional posterior probability density function of $\boldPsi$ is now replaced by 
\begin{align}
	\pi(\bvec{\Psi}|\cdot) \propto m(\bvec{\Psi}) \times
	|\mathbf{\Psi}|^{-(\tilde{\nu}_{\psi}+p+1)/2} \exp \left\{-\frac{1}{2}\tr \left(\mathbf{\Psi}^{-1}\tilde{\mathbf{S}}_\psi\right)\right\},
\end{align}
where
\begin{eqnarray}
m(\bvec{\Psi})
	&=&|\boldOmega_0|^{-1/2}\exp\left(-\frac{1}{2}(\boldh_1-\bvec{\mu})'\boldOmega_0^{-1} (\boldh_1-\bvec{\mu}) \right), \\
\tilde{\nu}_\psi 
	&=& \nu_\psi+T-1,\\
\nonumber
\tilde{\mathbf{S}}_\psi
	&= &  \mathbf{S}_\psi +\sum_{t=1}^{T-1}
	\left\{\boldh_{t+1} - \boldmu - \boldPhi(\boldh_t - \boldmu) - \boldLambda\boldR_t^{-1/2} \boldV_t^{-1/2} (\boldy_t - \boldm_t)\right\} \\
	& &\hspace{3cm}\left\{\boldh_{t+1} - \boldmu - \boldPhi(\boldh_t - \boldmu) - \boldLambda\boldR_t^{-1/2} \boldV_t^{-1/2} (\boldy_t - \boldm_t)\right\}'
	.
\end{eqnarray}
Thus we propose a candidate $\mathbf{\Psi}^{\dagger}$ from $\IW(\tilde{\nu}_{\psi},\tilde{\mathbf{S}}_{\psi})$, and accept it with probability $\min\{1, m(\mathbf{\Psi}^{\dagger})/m(\mathbf{\Psi})\}$.

\subsubsection{Generation of $\boldmu$}
We generate $\bvec{\mu}|\cdot \sim N(\bvec{\tilde{m}_\mu},\bvec{\tilde{\Psi}_\mu}),$
where
\begin{align*}
	&\tilde{\boldm}_\mu=\tilde{\boldPsi}_\mu
	\left[s_\mu^{-2}\bvec{m_\mu} +\boldOmega_0^{-1}\boldh_1
	+(\boldI-\bvec{\Phi})\bvec{\Psi}^{-1}\sum_{t=1}^{T-1}\left\{\boldh_{t+1}-\bvec{\Phi}\boldh_t
	- \boldLambda\boldR_t^{-1/2}\boldV_t^{-1/2}(\boldy_t - \boldm_t)\right\}
	\right] \\
	&\tilde{\boldPsi}_\mu= \left[s_\mu^{-2} \boldI+\boldOmega_0^{-1} + (T-1)(\boldI-\bvec{\Phi}) \bvec{\Psi}^{-1} (\boldI-\bvec{\Phi}) \right]^{-1}.
\end{align*}

\section{Proofs}
\subsection{Proof of Proposition 1}
\label{sec:proof1}
\noindent
{\bf Proof:}
\hspace{1mm}
Since $\boldR_{it}$ is positive definite, its principal submatrices are all positive definite. Further, noting that 
$\dtm{\boldR_t}=\dtm{\boldR_{it}}\times\dtm{1-\boldrho_{it}'\boldR_{it}^{-1}\boldrho_{it}}$,
the condition for $\mathbf{R}_t$ to be positive definite is 
$-\bvec{\rho}_{it}'\mathbf{R}_{it}^{-1}\bvec{\rho}_{it}>0$, 
which reduces to
\begin{align}
\label{eq:pd_ij}	-a_j\rho_{ij,t}^2-2\bm{b}_j'\boldrho_{i,-j,t}\rho_{ij,t}-\boldrho'_{i,-j,t}\mathbf{C}_j\boldrho_{i,-j,t} + 1 > 0
	,
\end{align}
Therefore the inequality (\ref{eq:pd_ij}) implies that the lower and upper bounds for $\rho_{ij,t}$ are given by (\ref{eq:pd}).
\begin{flushright}
$\Box$
\end{flushright}

\subsection{Proof of Proposition 2}
\label{sec:proof2}
We note that the probability density function 
for $\boldX \sim \Normal_{p,n}(\boldM, \boldPsi \otimes \boldSigma)$ is given by
\begin{align}
f(\boldX) = (2\pi)^{-np/2}|\boldPsi|^{-n/2}|\boldSigma|^{-p/2}
\times \exp
\left\{-\frac{1}{2}\tr
\left(\boldPsi^{-1} (\boldX - \boldM)\boldSigma^{-1} (\boldX - \boldM)'
\right)
\right\}.
\end{align}

\noindent
{\bf Proof:}
Since
\begin{eqnarray}
\lefteqn{f(\bm{h}_{t+1}|\bm{y}_t, \bm{g}_t, \bm{h}_t, \bm{m}_t,\bm{\theta})} &&
\nonumber
\\
	&\propto &  \abs{\boldPsi}^{-1/2}\exp\left\{
	-\frac{1}{2}\left(\boldh_{t+1} - \boldmu - \boldPhi(\boldh_t - \boldmu) - \boldLambda\boldR_t^{-1/2} \boldV_t^{-1/2} (\boldy_t - \boldm_t)\right)'\right. 
\nonumber
\\
	& & \hspace{2.5cm}\left.\boldPsi^{-1} \left(\boldh_{t+1} - \boldmu - \boldPhi(\boldh_t - \boldmu) - \boldLambda\boldR_t^{-1/2} \boldV_t^{-1/2} (\boldy_t - \boldm_t)\right)\right\},
\end{eqnarray}
and
\begin{eqnarray}
\pi(\mathbf{\Lambda}|\mathbf{\Psi}) & \propto &  
|\boldPsi|^{-p/2}|\mathbf{\Gamma}_0|^{-p/2}\exp
\left\{-\frac{1}{2}\tr
\left(\boldPsi^{-1} (\mathbf{\Lambda} - \boldM_0)\mathbf{\Gamma}_0^{-1} (\mathbf{\Lambda} - \boldM_0)'
\right)
\right\},
\end{eqnarray}
the conditional posterior probability density function of $\boldLambda$ is 
\begin{align}
\nonumber
\pi(\boldLambda \vert \cdot)
&\propto 
\exp\left[
-\frac{1}{2}
\left\{
\sum_{t=1}^{T-1}(\boldy_t - \boldm_t)'\boldV_t^{-1/2}\boldR_t^{-1/2'}\boldLambda'\boldPsi^{-1}\boldLambda\boldR_t^{-1/2}\boldV_t^{-1/2}(\boldy_t - \boldm_t)
\right. \right.
\\
\nonumber
&\quad\left. \left.
- 2 \sum_{t=1}^{T-1}(\boldh_{t+1} - \boldmu - \boldPhi(\boldh_t - \boldmu))'\boldPsi^{-1}\boldLambda\boldR_t^{-1/2}\boldV_t^{-1/2}(\boldy_t - \boldm_t)
\right\}
\right]
\times \pi(\boldLambda|\mathbf{\Psi}) 
\\
\nonumber
&\propto
\exp\left[
-\frac{1}{2}
\left\{
\tr(\boldPsi^{-1}\boldLambda(\boldA+\boldGamma_0^{-1})\boldLambda') - 2 \tr(\boldPsi^{-1}\boldLambda(\boldB +\boldGamma_0^{-1}\boldM_0) )
\right\}
\right] 
\\
&\propto
\exp
\left[
-\frac{1}{2}
\tr\left\{
\boldPsi^{-1}(\boldLambda - \boldM_1)(\boldA + \boldGamma_0^{-1})(\boldLambda - \boldM_1)'
\right\}
\right],
\end{align}
and the result follows.
\begin{flushright}
$\Box$
\end{flushright}
\subsection{Proof of Proposition 3}
\label{sec:proof3}
\noindent
{\bf Proof:} 
Using  $\boldLambda = \sum_{j=1}^q \bolde_j' \otimes \boldlambda_j$ where $\bolde_j$ is the $p\times 1$ unit vector with the $j$-th element being one,  the posterior probability density function of $\boldLambda$ is

\begin{align}
\nonumber
\lefteqn{\pi(\bm{\lambda} \vert \cdot)} & \\
\nonumber
&\propto
\exp\left[
-\frac{1}{2}
\left\{
	\sum_{t=1}^{T-1}\boldz_t' 
	\left(
	\sum_{j=1}^q\bolde_j\otimes \boldlambda_j'
	\right)
	\boldPsi^{-1}
	\left(
	\sum_{j=1}^q\bolde_j'\otimes \boldlambda_j
	\right)
	\boldz_t
	-2 \sum_{t=1}^{T-1}\boldeta_t'\boldPsi^{-1}
	\left(
	\sum_{j=1}^q\bolde_j'\otimes \boldlambda_j
	\right)
	\boldz_t 
\right\}
\right]
\\
\nonumber
& \hspace{2cm}\times\pi(\bm{\lambda})  
\\
\nonumber
&\propto \exp\left[
-\frac{1}{2}
\left\{
\sum_{t=1}^{T-1}
	\left(
	\sum_{j=1}^q\boldz_t'\bolde_j\otimes \boldlambda_j'
	\right)
	\boldPsi^{-1}
	\left(
	\sum_{j=1}^q\bolde_j'\boldz_t\otimes \boldlambda_j
	\right)
	-2 \sum_{t=1}^{T-1} \boldeta_t'\boldPsi^{-1}
	\left(
	\sum_{j=1}^q\bolde_j'\boldz_t\otimes \boldlambda_j
	\right)
\right\}
\right]
\\
\nonumber
& \hspace{2cm}\times\pi(\bm{\lambda})  
\\
\nonumber
&\propto \exp\left[
-\frac{1}{2}
\left\{
	\sum_{t=1}^{T-1}
	\sum_{i=1}^q\sum_{j=1}^q
 	(\bolde_i'\boldz_t \boldz_t'\bolde_j) \boldlambda_i'\boldPsi^{-1} \boldlambda_j
	-2 \sum_{t=1}^{T-1}\sum_{j=1}^q \bolde_j'\boldz_t \boldeta_t'\boldPsi^{-1}\boldlambda_j
\right\}
\right]\times\pi(\bm{\lambda}) 
\end{align}
\begin{align}
\nonumber
&\propto \exp\left[
-\frac{1}{2}
	\left\{
	\sum_{i=1}^q\sum_{j=1}^q
	 (\bolde_i'\boldA \bolde_j)
	\boldlambda_i'\boldPsi^{-1}\boldlambda_j- 2\sum_{j=1}^q\boldlambda_j'\boldPsi^{-1}\mathbf{B}'\bolde_j
	\right\}
	\right]\times\pi(\bm{\lambda}) \\
&\propto \exp\left[
-\frac{1}{2}
	\left\{
	\boldlambda'
	\left(\mathbf{\Gamma}_0^{-1}+\ \mathbf{A}_{1:q,1:q}\otimes \boldPsi^{-1}\right) \boldlambda
	- 2\boldlambda'
	\left(\mathbf{\Gamma}_0^{-1}\bm{m}_0
	+(\mathbf{I}_q\otimes\boldPsi^{-1}\mathbf{B}') \text{vec}\left(\{\bolde_1,\ldots,\bolde_q\}\right)
	\right)
	\right\}
	\right],
\end{align}
and the result follows.
\begin{flushright}
$\Box$
\end{flushright}
\footnotesize
\vspace{-5mm}
\bibliography{ref_MRSV_PRC_leverage_paper}	

\begin{thebibliography}{}

\bibitem[\protect\citeauthoryear{Andersen, Bollerslev, Diebold, and
  Ebens}{Andersen et~al.}{2001}]{AndersenBollerslevDieboldEbens(01)}
Andersen, T.~G., T.~Bollerslev, F.~X. Diebold, and H.~Ebens (2001).
\newblock The distribution of realized stock return volatility.
\newblock {\em Journal of Financial Economics\/}~{\em 61}, 43--76.

\bibitem[\protect\citeauthoryear{Andersen, Bollerslev, Diebold, and
  Labys}{Andersen et~al.}{2001}]{AndersenBollerslevDieboldLabys(01)}
Andersen, T.~G., T.~Bollerslev, F.~X. Diebold, and P.~Labys (2001).
\newblock The distribution of realized exchange rate volatility.
\newblock {\em Journal of the American Statistical Association\/}~{\em 96},
  42--55.

\bibitem[\protect\citeauthoryear{Barndorff-Nielsen and
  Shephard}{Barndorff-Nielsen and Shephard}{2002}]{BarndorffShephard(02)}
Barndorff-Nielsen, O.~E. and N.~Shephard (2002).
\newblock Econometric analysis of realised volatility and its use in estimating
  stochastic volatility models.
\newblock {\em Journal of the Royal Statistical Society: Series B (Statistical
  Methodology)\/}~{\em 64\/}(2), 253--280.

\bibitem[\protect\citeauthoryear{Barndorff-Nielsen and
  Shephard}{Barndorff-Nielsen and Shephard}{2004}]{BarndorffShephard(04)}
Barndorff-Nielsen, O.~E. and N.~Shephard (2004).
\newblock Econometric analysis of realized covariation: High frequency based
  covariance, regression, and correlation in financial economics.
\newblock {\em Econometrica\/}~{\em 72\/}(3), 885--925.

\bibitem[\protect\citeauthoryear{Chib, Nardari, and Shephard}{Chib
  et~al.}{2006}]{chib2006analysis}
Chib, S., F.~Nardari, and N.~Shephard (2006).
\newblock Analysis of high dimensional multivariate stochastic volatility
  models.
\newblock {\em Journal of Econometrics\/}~{\em 134\/}(2), 341--371.

\bibitem[\protect\citeauthoryear{Corsi}{Corsi}{2009}]{Corsi(09)}
Corsi, F. (2009).
\newblock A simple approximate long-memory model of realized volatility.
\newblock {\em Journal of Financial Econometrics\/}~{\em 7\/}(2), 174--196.

\bibitem[\protect\citeauthoryear{de~Jong and Shephard}{de~Jong and
  Shephard}{1995}]{deJongShephard(95)}
de~Jong, P. and N.~Shephard (1995).
\newblock The simulation smoother for time series models.
\newblock {\em Biometrika\/}~{\em 82}, 339--350.

\bibitem[\protect\citeauthoryear{Dobrev and Szerszen}{Dobrev and
  Szerszen}{2010}]{DobrevSzerszen(10)}
Dobrev, D.~P. and P.~J. Szerszen (2010).
\newblock The information content of high-frequency data for estimating equity
  return models and forecasting risk.
\newblock International Finance Discussion Papers, Board of Governors of the
  Federal Reserve System (U.S.).

\bibitem[\protect\citeauthoryear{Doornik}{Doornik}{2006}]{Doornik(06)}
Doornik, J. (2006).
\newblock {\em Ox: Object Oriented Matrix Programming}.
\newblock London: Timberlake Consultants Press.

\bibitem[\protect\citeauthoryear{Durbin and Koopman}{Durbin and
  Koopman}{2002}]{DurbinKoopman(02)}
Durbin, J. and S.~J. Koopman (2002).
\newblock Simple and efficient simulation smoother for state space time series
  analysis.
\newblock {\em Biometrika\/}~{\em 89}, 603--616.

\bibitem[\protect\citeauthoryear{Engle}{Engle}{2002}]{Engle(02)}
Engle, R. (2002).
\newblock Dynamic conditional correlation: A simple class of multivariate
  generalized autoregressive conditional heteroskedasticity models.
\newblock {\em Journal of Business \& Economic Statistics\/}~{\em 20-3},
  339--350.

\bibitem[\protect\citeauthoryear{Giacomini and White}{Giacomini and
  White}{2006}]{GiacominiWhite(06)}
Giacomini, R. and H.~White (2006).
\newblock Tests of conditional predictive ability.
\newblock ~{\em 74}, 1545--1578.

\bibitem[\protect\citeauthoryear{Han}{Han}{2006}]{Han2006}
Han, Y. (2006).
\newblock {Asset allocation with a high dimensional latent factor stochastic
  volatility model}.
\newblock {\em Review of Financial Studies\/}~{\em 19\/}(1998), 237--271.

\bibitem[\protect\citeauthoryear{Hansen, Huang, and Shek}{Hansen
  et~al.}{2012}]{HansenHuangShek(12)}
Hansen, P.~R., Z.~Huang, and H.~H. Shek (2012).
\newblock Realized \uppercase{GARCH}: A joint model for returns an realized
  measures of volatility.
\newblock {\em Journal of Applied Econometrics\/}~{\em 27\/}(6), 877--906.

\bibitem[\protect\citeauthoryear{Hansen, Lunde, and Voev}{Hansen
  et~al.}{2014}]{HansenLundeVoev(14)}
Hansen, P.~R., A.~Lunde, and V.~Voev (2014).
\newblock Realized beta {GARCH}: a multivariate {GARCH} model with realized
  measures of volatility.
\newblock {\em Journal of Applied Econometrics\/}~{\em 29\/}(5), 774--799.

\bibitem[\protect\citeauthoryear{Jin and Maheu}{Jin and
  Maheu}{2013}]{JinMaheu(13)}
Jin, X. and J.~M. Maheu (2013).
\newblock Modeling realized covariances and returns.
\newblock {\em Journal of Financial Econometrics\/}~{\em 11\/}(2), 335--369.

\bibitem[\protect\citeauthoryear{Jin and Maheu}{Jin and
  Maheu}{2016}]{JinMaheu(16)}
Jin, X. and J.~M. Maheu (2016).
\newblock Bayesian semiparametric modeling of realized covariance matrices.
\newblock {\em Journal Econometrics\/}~{\em 192\/}(1), 19--31.

\bibitem[\protect\citeauthoryear{Koopman and Scharth}{Koopman and
  Scharth}{2013}]{KoopmanScharth(13)}
Koopman, S.~J. and M.~Scharth (2013).
\newblock The analysis of stochastic volatility in the presence of daily
  realized measures.
\newblock {\em Journal of Financial Econometrics\/}~{\em 11}, 76--115.

\bibitem[\protect\citeauthoryear{Lopes and Carvalho}{Lopes and
  Carvalho}{2007}]{lopes2007factor}
Lopes, H.~F. and C.~M. Carvalho (2007).
\newblock Factor stochastic volatility with time varying loadings and {M}arkov
  switching regimes.
\newblock {\em Journal of Statistical Planning and Inference\/}~{\em
  137\/}(10), 3082--3091.

\bibitem[\protect\citeauthoryear{Lopes and West}{Lopes and
  West}{2004}]{lopes2004bayesian}
Lopes, H.~F. and M.~West (2004).
\newblock Bayesian model assessment in factor analysis.
\newblock {\em Statistica Sinica\/}~{\em 14\/}(1), 41--68.

\bibitem[\protect\citeauthoryear{Noureldin, Shephard, and Sheppard}{Noureldin
  et~al.}{2012}]{noureldin2012multivariate}
Noureldin, D., N.~Shephard, and K.~Sheppard (2012).
\newblock Multivariate high-frequency-based volatility (heavy) models.
\newblock {\em Journal of Applied Econometrics\/}~{\em 27\/}(6), 907--933.

\bibitem[\protect\citeauthoryear{Pitt and Shephard}{Pitt and
  Shephard}{1999}]{pitt1999time}
Pitt, M. and N.~Shephard (1999).
\newblock Time varying covariances: a factor stochastic volatility approach.
\newblock {\em Bayesian statistics\/}~{\em 6}, 547--570.

\bibitem[\protect\citeauthoryear{Shirota, Omori, Lopes, and Piao}{Shirota
  et~al.}{2017}]{ShirotaOmoriLopesPiao(17)}
Shirota, S., Y.~Omori, H.~F. Lopes, and H.~Piao (2017).
\newblock Cholesky realized stochastic volatility model.
\newblock {\em Econometrics and Statistics\/}~{\em 3}, 34--59.

\bibitem[\protect\citeauthoryear{So, Li, Asai, and Jiang}{So
  et~al.}{2016}]{SoLiAsaiJiang(16)}
So, M. K.~P., R.~W.~M. Li, M.~Asai, and Y.~Jiang (2016).
\newblock Stochastic multivariate mixture covariance model.
\newblock {\em Journal of Forecasting\/}.
\newblock in press.

\bibitem[\protect\citeauthoryear{Takahashi, Omori, and Watanabe}{Takahashi
  et~al.}{2009}]{TakahashiOmoriWatanabe(09)}
Takahashi, M., Y.~Omori, and T.~Watanabe (2009).
\newblock {Estimating stochastic volatility models using daily returns and
  realized volatility simultaneously}.
\newblock {\em Computational Statistics and Data Analysis\/}~{\em 53\/}(6),
  2404--2426.

\bibitem[\protect\citeauthoryear{Takahashi, Watanabe, and Omori}{Takahashi
  et~al.}{2016}]{TakahashiWatanabeOmori(16)}
Takahashi, M., T.~Watanabe, and Y.~Omori (2016).
\newblock Volatility and quantile forecasts by realized stochastic volatility
  models with generalized hyperbolic distribution.
\newblock {\em International Journal of Forecasting\/}~{\em 32\/}(2), 437--457.

\bibitem[\protect\citeauthoryear{Windle, Carvalho, et~al.}{Windle
  et~al.}{2014}]{WindleCarvalho(14)}
Windle, J., C.~M. Carvalho, et~al. (2014).
\newblock A tractable state-space model for symmetric positive-definite
  matrices.
\newblock {\em Bayesian Analysis\/}~{\em 9\/}(4), 759--792.

\bibitem[\protect\citeauthoryear{Zheng and Song}{Zheng and
  Song}{2014}]{ZhengSong(14)}
Zheng, T. and T.~Song (2014).
\newblock A realized stochastic volatility model with {Box-Cox} transformation.
\newblock {\em Journal of Business and Economic Statistics\/}~{\em 32\/}(4),
  593--605.

\end{thebibliography}
\end{document}